\documentclass[12pt]{iopart}

\usepackage[dvips]{graphicx}
\usepackage[]{cancel}
\usepackage[]{amssymb}
\usepackage[]{latexsym}
\usepackage[]{endnotes}

\newcommand{\bm}[1]{#1}

\newcommand{\eas}[0]{\begin{eqnarray*}}
\newcommand{\eae}[0]{\end{eqnarray*}}
\newcommand{\les}[0]{\begin{equation}}
\newcommand{\lee}[0]{\end{equation}}
\newcommand{\leas}[0]{\begin{eqnarray}}
\newcommand{\leae}[0]{\end{eqnarray}}
\newcommand{\mchss}[4]
{
\left\{
\begin{array}{cc}
#1 & #2   \\
#3 & #4
\end{array}
\right.
}

\newcommand{\mmat}[4]
{
\left[
\begin{array}{cc}
#1 & #2 \\
#3 & #4 
\end{array}
\right]
}

\newcommand{\matthree}[9]
{
\left[
\begin{array}{ccc}
#1 & #2 & #3 \\
#4 & #5 & #6 \\
#7 & #8 & #9 
\end{array}
\right]
}

\newcommand{\mvec}[2]
{
\left[
\begin{array}{c}
#1  \\
#2  
\end{array}
\right]
}

\newcommand{\mvecthree}[3]
{\small
\left[
\begin{array}{c}
#1  \\
#2  \\
#3  
\end{array}
\right]
}
\newcommand{\mvecfour}[4]
{
\left[
\begin{array}{c}
#1  \\
#2  \\
#3  \\
#4  
\end{array}
\right]
}
\newcommand{\mvecsix}[6]
{\small
\left[
\begin{array}{c}
#1  \\
#2  \\
#3  \\
#4  \\
#5  \\
#6  
\end{array}
\right]
}

\renewcommand{\footnote}[1]{\endnote}
\let\footnote=\endnote

\newcommand{\mynote}[1]{}

\begin{document}
\title[Chiral symmetry and its manifestation in optical responses in graphene]{Chiral symmetry and its manifestation in optical responses in graphene: interaction and multi-layers}


\author{Y. Hatsugai}
\address{
Institute of Physics, University of Tsukuba, Tsukuba 305-8571 Japan
}
\ead{hatsugai.yasuhiro.ge@u.tsukuba.ac.jp}

\author{T. Morimoto}
\address{
Condensed Matter Theory Laboratory, Riken, Saitama 351-0198, Japan
}

\author{T. Kawarabayashi}
\address{
Department of Physics, Toho University, Funabashi 274-8510, Japan
}

\author{Y. Hamamoto}
\address{
Institute of Physics, University of Tsukuba, Tsukuba 305-8571, Japan
}

\author{H. Aoki}
\address{
Department of Physics, University of Tokyo, Hongo, Tokyo 113-0033, Japan
}

\begin{abstract}
Chiral symmetry, 
fundamental in the physics of graphene, 
guarantees the existence of topologically stable doubled 
Dirac cones and anomalous behaviors of the 
zero-energy Landau level in magnetic fields.  
The crucial role is inherited 
in the optical responses and
many-body physics in graphene, which are explained 
in this paper.  
We also give an overview of multilayer 
graphene from the viewpoint of the 
optical properties and their relation with the 
chiral symmetry.  
\end{abstract}

\section{Introduction}
While the Dirac cone physics in graphene is forming a new branch 
in the condensed-matter physics, there is an increasing fascination with 
optical responses in graphene\cite{Nov05,Zhang05,Sadowski06}. 
The present article focuses on how the physics of massless Dirac cone 
and the optical properties of graphene are related.
  There, we stress the factor 
that links the two is the chiral symmetry.  
So the purpose of this paper is, first, to discuss the meaning 
of  the chiral symmetry in terms of fundamental ingredients  
in graphene including interaction and multi-layer systems as 
well as optical responses. 

So let us start with the Dirac cone, which comes from 
the symmetry of graphene.  
The honeycomb lattice structure of this material has 
two atoms per unit cell, and if we 
focus on the two $\pi$-bands arising from this, the system is effectively 
described by a traceless hermite 
Hamiltonian, $\bm{H}(\bm{k} )= \bm{R}(\bm{k} ) \cdot \bm{\sigma }$, 
which is parametrized by $\bm{R}$ with three real components.  This implies
that the Dirac cone has a co-dimension of three\cite{Berry84}. 
Since the momentum in two dimensions are described by two parameters, 
the Dirac cones do not arise in general, while 
the symmetry of the graphene at K and K' points gives rise 
to the Dirac cones. 
Here the key ingredient is the chiral symmetry of the honeycomb lattice.
When the lattice points are classified into two sublattices and the hopping 
of electrons is only allowed between them, 
the system is chiral-symmetric. Then 
we can show that the degree of freedom in $\bm{R}$ 
is reduced from three to two.  
This also ensures the topological stability 
(against various modifications) of the Dirac cones 
of the chiral symmetric
two-dimensional system, since the degeneracy cannot be removed 
by small but finite modifications as far as the chiral symmetry (or its 
extension) is respected. 

A conspicuous feature is graphene and related models is that 
the Dirac cones appear in pairs (at  K and K' 
points in graphene).  This is due to the well-known fermion doubling in 
chiral-symmetric systems, which is a two-dimensional analogue
of the Nielsen-Ninomiya theorem in four-dimensional lattice 
gauge theory\cite{Nielsen81,Hatsugai06gra}. 
In a real graphene, the chiral symmetry is not rigorously 
valid.  In the 
standard tight-binding parameters for graphene
\cite{dresselhaus-graphite02}, however, 
most of the chiral-symmetry-breaking parameters are significantly 
smaller than the chiral symmetric ones in magnitude.


\section{\bf Chiral symmetry}
Let us start with the chiral symmetry of fermions on a lattice, which 
is directly applicable to graphene. 
Assuming the lattice is bipartite, that is, 
all of the lattice points can be divided into two sublattice sites 
$\bullet  $ and $\circ $,  the non-interacting Hamiltonian 
${\cal H}_0$ is expressed as  block-off diagonal form,
\begin{eqnarray} 
{\cal H}_0 (\bm{D} )&=& {c} ^\dagger \bm{H}_0   {c} 
={c} _\bullet ^\dagger \bm{D}  {c} _\circ 
+{\rm h.c.},
\\
\bm{H} _0 &=& \mmat{\bm{O} }{\bm{D} }{\bm{D} ^\dagger  }{\bm{O} },
\quad c  = \mvec{{c}_\bullet } {{c}_\circ }, 
\quad {c} _\bullet = \mvecthree{c_{1\bullet }}{\vdots}{c_{N_\bullet }},
\quad {c} _\circ  = \mvecthree{c_{1 \circ  }}{\vdots}{c_{N_\circ }},
\end{eqnarray} 
where $N_\bullet $ and $N_\circ $ are the numbers
 of $\bullet $ and $\circ $ sites,
respectively, and
$\bm{D} $ is an $N_{\bullet }\times N_{\circ }$ matrix. 
We can introduce the chiral operator $\bm{\Gamma }$ that has
\begin{eqnarray}
\{\bm{H}_0, \bm{\Gamma }  \} &=& 0,
\ \
\bm{\Gamma }  = 
\mmat
{\bm{I}_{N_\bullet } }
{\bm{O}}
{\bm{O}}
{-\bm{I}_{N_\circ } }
\end{eqnarray} 
where
$
\bm{\Gamma } ^2 = \bm{I} _{N}
$,
$
N=N_\bullet + N_\circ  
$
and 
$
{\rm Tr}\,  \bm{\Gamma } = N_\bullet - N_\circ  
$.

The chiral symmetry $\{\bm{H}, \bm{\Gamma }\}  =0$ implies
that,
 if $\varphi_i$ is an eigenstate of $\bm{H} $ with energy $\lambda_i $
$\bm{\Gamma }\varphi_i $ is another eigenstate with energy $- \lambda_i $
($\bm{H}_0 \bm{\Gamma }  \varphi_i =
- \bm{\Gamma } \bm{H}_0  \varphi_i =
- \lambda _i \bm{\Gamma } \varphi_i $).  

Without loss of generality, let us assume $N_\bullet \ge N_\circ $. 
Then  the secular equation becomes 
\begin{eqnarray}
\fl 
0 = \det\nolimits_{N} (\lambda \bm{I}_N  -\bm{H}_0)= 
\det\nolimits_{N} \mmat
{\lambda \bm{I}_{N_\bullet }}{-\bm{D}}
{-\bm{D} ^\dagger }{\lambda \bm{I}_{N_\circ }}
=
\det\nolimits_{N} \mmat
{\lambda \bm{I}_{N_\bullet }}{-\bm{D}}
{O }{-\lambda ^{-1} \bm{D} ^\dagger \bm{D}+\lambda \bm{I}_{N_\circ }}
\\
=
\lambda ^{N_\bullet -N_\circ }\det\nolimits_{N_\circ }
(\lambda^2 \bm{I}_{N_{\circ }} -\bm{D} ^\dagger \bm{D}).
\end{eqnarray} 
This implies, at least,  $N_\bullet -N_\circ $ of the $\epsilon_{j} $'s 
are zero. 
These zero energy states (zero modes) are geometical
since 
the number of these exact zero-energy states is determined by the
geometrical structure of the lattice.

\subsection{\bf Chiral basis for the zero modes}
As for a zero-energy state, 
its chiral partner is degenerated in energy.
This implies that there is a gauge freedom in choosing the basis.
 In this paper, we focus on the zero-energy Landau level in graphene.  

Before proceeding, we have to take care a bit about 
what we really mean by the zero-energy Landau level. 
Rigorously speaking, the $n=0$ Landau level has 
the energy exactly equal to zero only 
for massless Dirac fermions in a continuous model. 
Conversely, if you consider a lattice model such as 
honeycom tight-binding model, the $n=0$ Landau level 
has in general a finite width, and the zero-energy condition 
is only realized in the limit of weak magnetic fields 
($\phi\to 0$ where $\phi$ is the flux per hexagon).  
In a numerical calculation for the honeycomb lattice model 
with a finite $\phi$, the $n=0$ Landau level has a 
small but finite width.   
Still, we want to project the problem onto the $n=0$ Landau level. 
For this purpose, 
we can define the $\epsilon$-zero mode by
collecting those states that have energies 
between $-\epsilon_{} $ and $\epsilon_{} $, 
where $\epsilon (>0)$ means the width of the Landau level. 
These naturally include the geometrical zero modes discussed above.

Now let us take an orthonormalized basis
for the  $\epsilon_{} $-zero modes
which form an $M$-dimensional multiplet
as
 $\bm{\psi} =(\psi_1,\cdots, \psi_M)$ ($\psi_ i ^\dagger \psi _j=\delta _{ij}$).
\begin{eqnarray}
\langle\psi_i^\dagger |\bm{H} _0^2| \psi_i \rangle 
 \le \epsilon ^2  
\end{eqnarray} 
for $^\forall i$.
This property is inherited by a 
unitary-transformed 
$\bm{\psi }_\omega = 
((\bm{\psi }_\omega )_1,\cdots,(\bm{\psi }_\omega )_M)= \bm{\psi} \bm{\omega }$
for $\omega \in U(M)$.
It imlies the $\epsilon$-zero modes are $U(M)$ gauge invariant. 

Then one can construct a normalized complete multiplet as
\begin{eqnarray}
\bm{\psi}_T &=& (\bm{\psi},\bm{\varphi} ,\bm{\Gamma } \bm{\varphi} ).
\end{eqnarray} 
where $\varphi$ is a multiplet of negative energy states 
$H_0 \varphi _i = -\lambda _i\varphi$, $\lambda _i> \epsilon_{} $
and $\Gamma \varphi $ is a multiplet of the
chiral partners of $\varphi$.
The normalization and the orthogonality of all the states are expressed as
\begin{eqnarray}
\bm{\psi}_T ^\dagger \bm{\psi}_T 
&=& 
\matthree
{\bm{\psi}^\dagger \bm{\psi}   }{\bm{\psi} ^\dagger \bm{\varphi}  }{\bm{\psi}  ^\dagger \bm{\Gamma } \bm{\varphi}  }
{\bm{\varphi}^\dagger \bm{\psi}  }{\bm{\varphi} ^\dagger \bm{\varphi}  }{\bm{\varphi} ^\dagger \bm{\Gamma } \bm{\varphi}  }
{ \bm{\varphi}^\dagger \bm{\Gamma }\bm{\psi}  }{ \bm{\varphi} ^\dagger\bm{\Gamma } \bm{\varphi}  }{ \bm{\varphi} ^\dagger  \bm{\varphi}  }
=
\matthree
{I_M}{0}{0}
{0}{I_{M'}}{0}
{0}{0}{I_{M'}}
\label{normorth}
\end{eqnarray} 
where $M'$ is the dimension of the negative (positive) enerygy 
multiplet $\varphi$ ($\Gamma \varphi$).
The completeness reads
\begin{eqnarray} 
\bm{\psi}_T \bm{\psi}_T ^\dagger 
&=&
 \bm{\psi} \bm{\psi} ^\dagger 
+ \bm{\varphi} \bm{\varphi } ^\dagger 
+ \bm{\Gamma } \bm{\varphi} \bm{\varphi } ^\dagger \bm{\Gamma }  = \bm{I} _{N}.
\end{eqnarray} 

With this multiplet $\bm{\psi}_T $, 
 $\bm{H}_0 $ is block diagonalized as
\begin{eqnarray}
\bm{H} _0 \bm{ \psi}_T = \bm{\psi} _T 
\matthree
{{\cal E}}{0}{0}
{0}{-\Lambda}{0}
{0}{0}{\Lambda}
\\
{\cal E}
 = \bm{\psi} ^\dagger  {\bm{H}_0 } \bm{\psi} ,\quad 
\bar{\cal E}^2 \equiv  {\rm Tr} {\cal E}^2/M\le \epsilon^2,\quad
{\Lambda} = {\rm diag}\, ( \lambda_1,\cdots, \lambda _{M'}) \
(\lambda_{j}> \epsilon_{} ).
\end{eqnarray}

Due to the completeness and orthogonality relations 
Eq.(\ref{normorth}),  we have
\begin{eqnarray}
\bm{\Gamma } \bm{\psi}  &=&
\bm{\psi}_T \bm{\psi}_T ^\dagger    \bm{\Gamma } \bm{\psi}  =
 \bm{\psi} \bm{\psi} ^\dagger    \bm{\Gamma } \bm{\psi} 
+ \bm{\varphi} \bm{\varphi } ^\dagger    \bm{\Gamma } \bm{\psi} 
+ \bm{\Gamma } \bm{\varphi} \bm{\varphi } ^\dagger \bm{\Gamma } ^2  \bm{\psi} 
=
\bm{\psi} \bm{\Gamma } _0, 
\\
\bm{\Gamma } _0 &=& \bm{\psi} ^\dagger \bm{\Gamma } \bm{\psi}, 
\end{eqnarray} 
where the
$M\times M$-dimensional hermitian  matrix $\bm{\Gamma }_0  $  satisfies
$
\bm{\Gamma } _0 ^2 = \bm{I} _M
$.
Its trace is evaluated as 
\begin{eqnarray} 
{\rm Tr} \bm{\Gamma } _0 = 
{\rm Tr} \bm{\Gamma } (\bm{\psi} \bm{\psi} ^\dagger )
= {\rm Tr} \, \bm{\Gamma } 
- 2{\rm Tr}\, \bm{\Gamma } (\bm{\varphi} \bm{\varphi} ^\dagger )
= {\rm Tr}\,  \bm{\Gamma } =N_\bullet-N_\circ 
\end{eqnarray} 
Then $\bm{\Gamma } _0$ is diagonalized as
\begin{eqnarray}
\bm{\Gamma } _0 = \bm{\omega }_\Gamma 
\bm{\Gamma }_{M_+M_-} 
 \bm{\omega }_\Gamma   ^\dagger ,
\\
\bm{\Gamma }_{M_+M_-} = \mmat{\bm{I} _{M_+}}{O}{O}{-\bm{I} _{M_-}},
\quad M_+-M_- = N_\bullet -N_\circ ,
\end{eqnarray} 
where $\bm{\omega }_\Gamma \in U(M) $ with $M_++M_-=M$.
Now let us define
a chiral multiplet $\bm{\psi}_\Gamma =\bm{\psi} \bm{\omega }_\Gamma   $
that satisfies
\begin{eqnarray}
\bm{\Gamma } \bm{\psi} _\Gamma = 
\bm{\psi}_\Gamma   \bm{\Gamma }_{M_+M_-},\quad 
\bm{\psi} _\Gamma = (\bm{\psi}_+,\bm{\psi}_-  ), 
\\
\bm{\Gamma } \bm{\psi} _+ = \bm{\psi} _+, \quad
\bm{\psi} _+ = (\psi_{1+},\cdots, \psi_{M_++}), 
\\
\bm{\Gamma } \bm{\psi} _- = -\bm{\psi} _-. 
\quad
\bm{\psi} _- =  (\psi_{1-},\cdots, \psi_{M_--}).
\end{eqnarray}
Namely, $\psi_{i\pm }$'s  are eigenstates of $\bm{\Gamma } $ as
\begin{eqnarray}
\bm{\Gamma } \psi_{i\pm}= \pm \psi_{i\pm},
\end{eqnarray} 
\begin{eqnarray}
\bm{\psi_+} = 
\mvec
{\bm{\psi}_\bullet }{\bm{0} },\quad
\bm{\psi_-} = 
\mvec
{\bm{0} }{\bm{\psi}_\circ  },
\end{eqnarray} 
where $\bm{\psi}_\bullet $ is an $N_\bullet \times M_+$ matrix and 
 $\bm{\psi}_\circ  $ an $N_\circ \times M_-$ matrices.



\section{\bf Electron-electron interaction 
and chiral condensate 
in a
magnetic field}

Let us move on to discuss the electron-electron interaction 
in graphene in magnetic fields. 
\subsection{\bf Two-body interactions}
As for the electron-electron interaction, we take the 
two-body interaction $V_{ij}n_i n_j$ between sites
$\langle i j  \rangle $, which can include long-range interactions. 
Note that 
 \begin{eqnarray} 
 n_{i    } n_{j   } = 
 c_{i  } ^\dagger c_{i  } c_{j  } ^\dagger c_{j  }
= 
 c_{i  } ^\dagger c_{j  } ^\dagger c_{j  }c_{i  } 
= 
-1+ n_{i  } +n_{j  }
+ c_{i  }c_{j  } c_{j  } ^\dagger c_{i  } ^\dagger .
\end{eqnarray} 
We can express the interaction in an 
electron-hole symmetric form, at half-filling, as
 \begin{eqnarray}
\fl\quad 
{\cal H}_{int} &=& \sum_{ i j  \rangle} V_{ij}
\left(  n_i- \frac {1}{2} \right) \left(  n_j- \frac {1}{2} \right)
=  \frac {1}{2} \sum_{\langle ij \rangle} V_{ij}
\bigg[
 c_i ^\dagger c_j  ^\dagger c_j c_i 
+
(c\rightleftarrows c ^\dagger )
\bigg]+{\rm const}.
\end{eqnarray}

\subsection{\bf Ground state within the projected subspace}
We assume that the electron-electron interaction is sufficiently 
weak as compared with
the Landau gap between the $n=0$ Landau level and those for 
$n=\pm 1$. 
When the filling factor of the $n=0$ LL is $\nu<1$, 
the  ground state  is then 
given by the configurations of the $n=0$ LL states, 
while the filled states below them can be regarded as
the ``Dirac sea". 
The one-particle spectrum in graphene
is given by the $\epsilon_{} $-zero modes and the 
chiral pairs of  non-zero energy states as discussed above. 
Then the kinetic energy from  the $\epsilon_{} $-zero modes
is negligible, 
so that the ground state is given peturbatively (i.e., in the interaction energy 
that is smaller than the Landau gap) as 
\begin{eqnarray} 
|\Psi \rangle  &=& 
\sum_{i_1,\cdots, i_M\subset\{1,\cdots, M\}}
C_{i_1,\cdots, i_M} d_{i_1} ^\dagger \cdots d_{i_M} ^\dagger 
| D_< \rangle ,
\ \
| D_< \rangle = \prod_{j=1}^{M'} {d_<}_{j} ^\dagger | 0 \rangle ,
\end{eqnarray} 
where $|D_< \rangle $ is the filled Dirac sea 
( $d_i | D_< \rangle =0$ for all $i$'s), 
$\{i_1,\cdots, i_M\}$ a subset of the $\epsilon_{} $-zero modes, and
$M$ the number of particles. 
The free-fermion Hamiltonian ${\cal H}_0$
(i.e., the kinetic energy) is written as
\begin{eqnarray}
{\cal H}_0 = 
c ^\dagger \bm{\psi_T}
\,{\rm diag}\, ({\cal E},- \bm{\Lambda } ,\bm{\Lambda } )\bm{\psi}_T ^\dagger  c
= 
{d}^\dagger {\cal E} d 
-{d_<}^\dagger \bm{\Lambda } d_<
+{d_>}^\dagger \bm{\Lambda } d_>,
\\
\mvecthree{d }{d_<}{d_>} =\bm{\psi}  _T ^\dagger c =
\mvecthree{\psi ^\dagger c }{\bm{\varphi }^\dagger  c }{\bm{\varphi}^\dagger \bm{\Gamma } c  },  
\ \
c =
\bm{\psi} _T\mvecthree{d }{d_<}{d_>}
=(\bm{\psi} ,\bm{\varphi},\bm{\Gamma } \bm{\varphi} )
\mvecthree{d }{d_<}{d_>}
\end{eqnarray} 
where 
$d ^\dagger =(d_1 ^\dagger ,\cdots, d_M ), $
 ${d_<} ^\dagger =({d_<}_1 ^\dagger ,\cdots, {d_<}_{M'} ^\dagger ) $, 
and
 ${d_>} ^\dagger =({d_>}_1 ^\dagger ,\cdots, {d_>}_{M'} ^\dagger ) $
are standard fermion creation operators 
for
the $\epsilon_{} $-zero mode, the negative energies and positive ones 
as
 $\{d_\ell, d_{\ell'} ^\dagger \}= \delta _{\ell\ell'}$,
 $\{d_{<\ell}, d_{<\ell'} ^\dagger \}= \delta _{\ell\ell'}$,
 $\{d_{>\ell}, d_{>\ell'} ^\dagger \}= \delta _{\ell\ell'}$, etc.

\subsection{\bf Projected interaction and its symmetry}
Now let us take the chiral basis for the $\epsilon_{} $-zero modes
to describe the interaction between the particles. 
We assume that the perturbative ground state 
within  the configuration of the $n=0$ Landau level 
is determined by the projected interaction $\widetilde{\cal H}_{int}$ .
The projected Hamiltonian is defined as
\begin{eqnarray}
\fl\quad 
\widetilde{\cal H}_{int}
= 
 \frac {1}{2} \sum_{ij}V_{ij}
[ \tilde {c}_{i } ^\dagger  \tilde {c}_{j } ^\dagger 
 \tilde {c}_{j } \tilde {c}_{i } 
+
(c\rightleftarrows c ^\dagger )]
=
 \frac {1}{2} \sum_{ij}V_{ij}
(d ^\dagger \bm{\psi}^\dagger  )_i
(d ^\dagger \bm{\psi}^\dagger  )_j
(\bm{\psi}  d)_j
(\bm{\psi}  d)_i + 
C.c.,
\end{eqnarray}
where 
$C.c.$ is a charge conjugation defined below
and $ \tilde{c}_{i}$ is a  projected fermion defined as
\begin{eqnarray}
\tilde{c} \equiv 
\bm{\Psi} _T\mvecthree{d }{0}{0}
=(\bm{\psi} ,\bm{\varphi},\bm{\Gamma } \bm{\varphi} )
\mvecthree{d }{0}{0}=\bm{\psi}  d
=\bm{\psi}  
\bm{\psi} ^\dagger c = \bm{P} c,
\end{eqnarray}
with $\bm{P} =\bm{\psi} \bm{\psi}  ^\dagger=\bm{P}^\dagger   $ 
being the projection to the 
$\epsilon_{} $-zero modes.
These $\widetilde {c}_i $'s 
are not canonical fermions, since
\begin{eqnarray}
\{{\tilde{c}}_i  ,\tilde{c}_j ^\dagger \} = 
\{
(\bm{P} )_{ii'}c_{i'}
,
c_{j'} ^\dagger (\bm{P}^\dagger  )_{j'j}
\}
=(\bm{P} \bm{P} ^\dagger   )_{ij}
=(\bm{P}^2 )_{ij}=(\bm{P} )_{ij}.
\end{eqnarray} 
Note that the projected Hamiltonian 
is clearly positive semi-definite.

\subsection{\bf  Chiral transformation  and charge conjugation}
Let us take a chiral basis to define  canonical fermions $d_i$ for the
$\epsilon_{} $-zero modes, defined as
\begin{eqnarray}
{d}  = \mvec{{d}_+ }{{d}_- }= \bm{\psi}_\Gamma ^\dagger  c
=\mmat
{\bm{\psi}_\bullet ^\dagger  }{\bm{O} }
{\bm{O} }{\bm{\psi}_\circ  ^\dagger  } c
=
\mvec
{\bm{\psi}_ \bullet ^\dagger c_\bullet  }
{\bm{\psi}_ \circ  ^\dagger c_\circ   }
\\
\ \
{d}_+ = \mvecthree{d_{1+}}{\vdots}{d_{M_++}},
\ \
{d}_-= \mvecthree{d_{1-}}{\vdots}{d_{M_--}}
\\
\tilde{c} = 
\mvec
{ \widetilde {c} _\bullet }
{ \widetilde {c} _\circ }
=
\mmat
{\bm{\psi}_\bullet  }{\bm{0}  }
{\bm{0}  }{\bm{\psi}_\circ  }
d
=  
\mvec
{\bm{\psi}_\bullet {d}_+} 
{\bm{\psi}_\circ   {d}_-} 
\end{eqnarray}
We can define a chiral transformation  
${\cal U}_\theta$ as
\begin{eqnarray}
 {\cal O} \mapsto
{\cal U}_\theta ^\dagger  {\cal O}{\cal U}_\theta,
\quad 
{\cal U}_\theta = e ^{i\theta {\cal G}},
\end{eqnarray}
where the generator of the transformation, 
the chirality ${\cal G}$,  is given as
\begin{eqnarray} 
{\cal G} = \tilde{c}^\dagger  \bm{\Gamma } \tilde{c} = 
\sum_{i\in \bullet }\tilde {c} _i ^\dagger \tilde {c} _i
-
\sum_{i\in \circ  }\tilde {c} _i ^\dagger \tilde {c} _i
=
{d}^\dagger \bm{\psi}_\Gamma ^\dagger   \bm{\Gamma } \bm{\psi}_\Gamma {d} = 
{d}^\dagger \bm{\Gamma }_0  {d} = 
d_
+ ^\dagger d_+-d_- ^\dagger d_-.
\end{eqnarray} 
The chiral transformation
operates as
\begin{eqnarray} 
 d \mapsto {d}_\theta = 
{\cal U}_\theta ^\dagger  
\, {d} \,
{\cal U}_\theta 
=
e^{ i \theta {\cal G}}
\,  {d} \,
e^{ -i \theta {\cal G}}
= e^{-i\theta \bm{\Gamma }_0}
{d} 
 =\mvec{e^{-i\theta }{d}_+ }{e^{i\theta}{d}_- } 
\\
\tilde {c}_i  \mapsto 
\mchss
{e^{-i\theta }\tilde {c}_i }{i\in \bullet }
{e^{+i\theta }\tilde {c}_i }{i\in \circ  }
\end{eqnarray} 
Then it is clear that the projected interaction is invariant under the
chiral transformation as
\begin{eqnarray}
\widetilde{\cal H}_{int} \mapsto e^{-i\theta  {\cal G}} 
\widetilde{\cal H}_{int}
e^{+i\theta {\cal G}}
&=& \widetilde{\cal H}_{int}.
\end{eqnarray} 
For an infinitesimal transformation 
this implies a conservation law, 
\begin{eqnarray}
[{\cal G}, \widetilde{\cal H}_{int}] &=& 0.
\end{eqnarray} 

Let us next define a charge conjugation which is 
 anti-unitary as 
\begin{eqnarray}
 O\mapsto 
 {\cal A}_C ^\dagger   O {\cal A}_C,
\\
{\cal A}_C =
K \prod_{\ell=1}^M
(d_{\ell}  +d_{\ell}  ^\dagger )
 \prod_{\ell=1}^{M'}
(d_{<\ell}  +{d}_{<\ell}  ^\dagger )
(d_{>\ell}  +{d}_{>\ell}  ^\dagger ),
\end{eqnarray}
where $K$ is complex-conjugation with $ K^2=1$. 
Its operation is, for example,  
\begin{eqnarray}
{\cal A}_C ^\dagger \, d_\ell \, {\cal A}_C =(-)^{N-1} d_\ell  ^\dagger,\quad 
{\cal A}_C ^\dagger \, d_\ell ^\dagger \, {\cal A}_C =(-)^{N-1} d_\ell.
\end{eqnarray} 
Then the invariance of the interaction Hamiltonian 
under the charge conjugation,
\begin{eqnarray}
{\cal A}_C ^\dagger \widetilde{\cal H}_{int} {\cal A}_C = \widetilde{\cal H}_{int}
\end{eqnarray} 
 trivially follows.

\subsection{\bf Chiral condensate as doubly-degenerate ground states}
In this section, we restrict ourselves to consider a repulsive interaction 
that only acts between $\bullet $ and $\circ $ sites.  
The projected interaction in this case is written as
\begin{eqnarray} 
\fl\qquad \widetilde{\cal H}_{int}= 
 \frac {1}{2} \sum_{ i_\bullet j_\circ  } V_{ i_\bullet j_\circ  }
(
 \tilde {c}_{i \bullet} ^\dagger  \tilde {c}_{j \circ} ^\dagger 
 \tilde {c}_{j \circ} \tilde {c}_{i \bullet} 
+
C.c)
\\
\fl\qquad
= \frac {1}{2} \sum_{ i_\bullet j_\circ}
V_{i_\bullet j_\circ  }
\bigg[
( {d}_+ ^\dagger \bm{\psi}_\bullet  ^\dagger )_i
( {d}_- ^\dagger \bm{\psi}_\circ  ^\dagger )_j
(  \bm{\psi}_\circ  {d}_- )_j
(  \bm{\psi}_\bullet  {d}_+  )_i
+
(  \bm{\psi}_\bullet  {d}_+  )_i
(  \bm{\psi}_\circ  {d}_- )_j
( {d}_- ^\dagger \bm{\psi}_\circ  ^\dagger )_j
( {d}_+ ^\dagger \bm{\psi}_\bullet  ^\dagger )_i
\bigg],
\end{eqnarray} 
where $^\exists V_{ i_\bullet j_\circ  }> 0$.
Here we allow any repulsive interaction between the
 $\bullet $ and $\circ $ sites.  

Now we can readily see that the two states, 
\begin{eqnarray}
\widetilde{\cal H}_{int} |G_\pm \rangle = 0,
\quad\quad
|G _\pm \rangle = d_{1\pm} ^\dagger \cdots d_{M_\pm \pm} ^\dagger | D_< \rangle , 
\end{eqnarray} 
which have maximum and minimum chiralities, respectively, 
are special, since each of them  has vanishing interaction energy.
As for $|G_+ \rangle $, for example, 
 all $d_{\ell+}$ states are filled, 
$( {d}_+ ^\dagger \bm{\psi}_\bullet  ^\dagger )_i |G_+ \rangle =0$.
Also none of the   $d_{\ell -}$ states are occupied, 
$(  \bm{\psi}_\circ  {d}_- )_j|G_+ \rangle =0$.
Since the interaction Hamiltonian $\widetilde {\cal H}_{int}$ is 
semi-positive definite, these two states are 
ground states of $\widetilde {\cal H}_{int}$ with a degeneracy of two .

As for the chiral transformation, we have
\begin{eqnarray}
 |G_\pm  \rangle \mapsto {\cal U}_\theta ^\dagger  |G_\pm  \rangle = 
e^{\mp i M_\pm\theta} | G_\pm \rangle , 
\end{eqnarray} 
where ${\cal U}_\chi$ is the chiral operation defined above.  
This property follows, since $ |G_\pm  \rangle$ 
are eigenstates of the chirality ${\cal G}$ as 
\begin{eqnarray}
{\cal G} | G_\pm \rangle = \pm M_\pm | G_\pm \rangle.
\end{eqnarray} 
Thus we have shown that the ground states 
are the {\it chiral condensates}, 
where the chirality is macroscopic ($M_\pm \sim N$). 

As for the charge conjugation, it operates as
\begin{eqnarray} 
 | G_\pm \rangle \mapsto 
(-)^S d_{1\mp} ^\dagger \cdots d_{M_\mp \mp} ^\dagger | D_> \rangle, 
\end{eqnarray}
where the $|D_> \rangle $ is the positive Dirac sea up to the sign. 
 
\subsection{\bf Chiral-condensate doublet as a Hall insulator}
The Hall conductance of the doubly-degenerate
chiral condensates can be calculated with the Niu-Thouless-Wu formula as 
\begin{eqnarray} 
\sigma _{xy} = \frac {e^2}{h} \frac {1}{N_D}  C, \quad 
C = \frac {1}{2\pi i} \int {\rm Tr}_2\, F,\quad
F = dA + A^2,\quad
A = \Psi ^\dagger d \Psi,
\end{eqnarray}
where 
$\Psi = (|G_+ \rangle , |G_- \rangle ) $
is the chiral-condensed ground states with the degeneracy of  
$N_D=2$. 

The doublet, being degenerate, can be mixed, but diagonalization of the
chirality ${\cal G}$ acts to fix the gauge.
Here we assume a finite energy gap above the ground state multiplet. 
Then the Chern number of the doublet becomes well-defined, 
and given by the sum of each as
\begin{eqnarray}
C &=& C_++C_-,
\end{eqnarray} 
where $C_\pm$ is the Chern number for $| G_\pm \rangle $.
Further, we can decompose the condensate as
\begin{eqnarray}
C_\pm &=& C_{\psi_\pm }+C_{D_<},
\end{eqnarray}
\begin{eqnarray} 
C_{\psi_\pm } &=& \frac {1}{2\pi i } \int
 {\rm Tr}_{M_\pm} d \bm{\psi}_\pm ^\dagger d \bm{\psi}_\pm,
\quad
C_{D_<}=\frac {1}{2\pi i } \int  \langle d D_< | d D_< \rangle , 
\end{eqnarray} 
where $C_{D_<}$ is the Chern number of the filled Dirac sea.
Since the chiral operator $\Gamma $ is unitary,
the Chern number of the positive Dirac sea is the same as that of
the negative Dirac sea as
\begin{eqnarray}
C_{D_>}=\frac {1}{2\pi i } \int  \langle d D_> | d D_> \rangle =
C_{D_<}\equiv C_{D}.
\end{eqnarray}  
Since the charge conjugation is anti-unitary,  it implies
\begin{eqnarray} 
C_{D_>} + C_{\psi_-} = -(C_{D_<} + C_{\psi_+} ),
\end{eqnarray} 
which in turn gives the total Chern number of the doubly degenerate
chiral condensate as 
\begin{eqnarray}
C &=& \big(C_{D_<} + C_{\psi_-} \big)+
\big(C_{D_<} + C_{\psi_+} \big)=
 C_{\psi_-} + C_{\psi_+} + 2C_D =0.
\end{eqnarray} 
This implies that the half-filled state 
formed by the chiral doublet 
 is a {\it Hall insulator} with the topological degeneracy of 2.

\section{Multilayer graphene and chiral symmetry}

\subsection{Bilayer/trilyer Hamiltonians}

\begin{figure}
\begin{center}
\includegraphics[width=6.0cm]{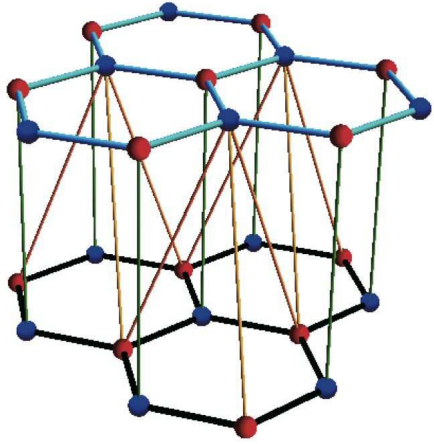}
\end{center}
\caption{
Lattice structure of the  bilayer graphene.
Interlayer couplings $\gamma _1$ and $\gamma _3$ are denoted
by green and yellow bonds, respectively.
}
\label{fig:unit}
\end{figure}

Let us start with the bilayer graphene 
in magnetic fields\cite{Hatsugai06gra}.  
With $ \bm{j}  =(j_1,j_2)$ denoting two-dimensional coordinates, 
$\bm{e}_{1,2} $ the unit translations, and 
$\varphi$ the total flux per hexagon in units of the  
flux quantum $\varphi_0=h/e$ (see Fig.\ref{fig:unit}), 
the Hamiltonian is written as
\begin{eqnarray}
  \fl\qquad
{\cal H}^{\rm bilayer} = H_d + H_u + H_{ud},
\\
\fl\qquad
{\cal H}_d = 
 \gamma _0\sum_j\bigg[
 d_\bullet ^\dagger  ( \bm{j} ) d_\circ (\bm{j} ) 
\nonumber \\
\qquad+
 e^{i2\pi \varphi j_1 }
 d_\bullet ^\dagger  (\bm{j} ) d_\circ (\bm{j} -\bm{e} _2) 
 +
 e^{-i2\pi \varphi (j_1 + \frac {3}{6} )} d_\bullet ^\dagger  (\bm{j} +\bm{e} _1) d_\circ (\bm{j} )  
+ {\rm h.c.}
\bigg],
\\
\fl\qquad
{\cal H}_u = {\cal H}_d (d\to u,j_1\to j_1+2/6)
\\
\fl\qquad
{\cal H}_{ud}=  
\sum_j\bigg[ 
\gamma _1 
u_\bullet   ^\dagger (\bm{j} ) d_ \circ (\bm{j} )
\nonumber \\
\fl\quad
+ \gamma _3
 \bigg(
 e^{i2\pi \varphi (j_1 +\frac {1}{6} )} d_\bullet  ^\dagger   (\bm{j})
+
d_\bullet ^\dagger  (\bm{j}+ \bm{e}_1 )
+
 e^{-i2\pi \varphi (j_1 +\frac {4}{6} )}d_\bullet ^\dagger 
 (\bm{j}+ \bm{e}_1 -\bm{e}_2 )
\bigg)
u_ \circ   (\bm{j}-\bm{e}_2 )
+ {\rm h.c.}
\bigg]
\end{eqnarray} 
Here "$u (d)"$ denote up (down) layers, 
$\gamma _0$ is the nearest-neighbor hopping within each layer, and 
$\gamma _1$ the inter-layer hopping perpendicular to the
layer.  
The hopping $\gamma _3$, the second largest inter-layer hopping, 
is along an oblique direction, causes
trigonal deformation of the energy dispersion (trigonal warping)\cite{Dress81} 
in zero magnetic field.  In a finite magnetic field, 
the hopping acquires Peierls phases appearing 
in the Hamiltonian above, again due to the 
oblique directions of the hopping.

As for the trilayer, we have
\begin{eqnarray}
\fl\qquad 
{\cal H}^{\rm trilayer} = {\cal H}_d + {\cal H}_u +{\cal H}_t
 + {\cal H}_{ud} + {\cal H}_{ut}^{ABA, ABC}
\\
\fl\qquad
{\cal H}_t = {\cal H}_d(d\to t,j_1\to j_1+4/6)
\\
\fl\qquad H_{ut}^{ABC}
={\cal H}_{ud}(d\to u, u\to t, j_1\to j_1+2/6)
\\
\fl\qquad H_{ut}^{ABA}: H_{ud} (d\to t, u \to u). 
\end{eqnarray} 
Here we have added "$t"$ (top layer) on top of 
the "$d, u"$ (down, up) layers, and ABA (ABC) stand for the trilayer graphene with the ABA (ABC) stacking.

Now let us Fourier-transform the fermion operators, first 
along the $\bm{e}_2 $ direction as
\begin{eqnarray}
\fl\qquad 
d_\alpha,u_\alpha,t_\alpha (j ) =
 \int _{-\pi }^{\pi} \frac {dk_2}{2\pi} 
e^{ik_2 j_2} d_\alpha,u_\alpha,t_\alpha (j_1,k_2).
\end{eqnarray}
Then we have
\begin{eqnarray}
\fl\qquad 
{\cal H}^{b} = \int_{-\pi}^\pi \frac {dk_2}{2\pi} {\cal H}^{ 1D,b}(k_2),
\qquad 
{\cal H}^{t} = \int_{-\pi}^\pi \frac {dk_2}{2\pi} {\cal H}^{ 1D,t}(k_2),
\\
\fl\qquad 
{\cal H} ^{ 1D,b}(k_2)
=
{\cal H}_d^{ 1D}(k_2)
+
{\cal H}_u^{ 1D}(k_2)
+
{\cal H}_{du}^{ 1D}(k_2),
\\
\fl\qquad 
{\cal H} ^{ 1D,t}(k_2)
=
{\cal H}_d^{ 1D}(k_2)
+
{\cal H}_u^{ 1D}(k_2)
+
{\cal H}_t^{ 1D}(k_2)
+
{\cal H}_{du}^{ 1D}(k_2)
+
{\cal H}_{ut}^{ 1D,ABC(ABA)}(k_2),
\\
\fl\qquad 
{\cal H}_d^{ 1D}(k_2) =
  \gamma _0\sum_{j_1}\bigg[
\big(
1 
+
 e^{i(2\pi \varphi j_1 -k_2)}
\big)
 d_\bullet ^\dagger  (j_1,k_2 ) d_\circ (j_1,k_2) 
\nonumber \\
+
e^{-i 2\pi\varphi(j_1+ \frac {3}{6} )} d_\bullet ^\dagger  (j_1 +1,k_2) d_\circ (j_1 ,k_2)  
+ {\rm h.c.}
\bigg],
\\
\fl\qquad 
{\cal H}_u^{ 1D}(k_2) =
{\cal H}_d^{ 1D}(k_2) : d\to u, j_1\to j_1+2/6,
\\
\fl\qquad 
{\cal H}_t^{ 1D}(k_2) =
{\cal H}_d^{ 1D}(k_2) : d\to t, j_1\to j_1+4/6,
\\
\fl\qquad 
{\cal H}_{du}^{\rm 1D}(k_2) =
\sum_{j_1}\bigg[ 
\gamma _1 
u _ \bullet ^\dagger (j_1,k_2 ) d_ \circ (j_1,k_2 )
\nonumber \\
\fl\
+ \gamma _3 
 \bigg(
 e^{i(-k_2+2\pi \varphi (j_1 +\frac {1}{6} ))} d_\bullet ^\dagger 
  (j_1,k_2)
+
(
e^{-k_2}
+
 e^{-i 2\pi \varphi (j_1+ \frac {4}{6})  }
)
d_\bullet ^\dagger  (j_1+1,k_2)
\bigg) 
u_ \circ   (j_1,k_2)
+ {\rm h.c.}
\bigg],
\\
\fl\qquad 
{\cal H}_{ut}^{ 1D,ABC}(k_2) 
={\cal H}_{du}^{1D}(k_2) : d\to u,u\to t,j_1\to j_1+2/6,
\\
\fl\qquad  
{\cal H}_{ut}^{\rm 1D,ABA}(k_2) 
={\cal H}_{du}^{1D}(k_2) : d\to t,u\to u.
\end{eqnarray}

\begin{figure}
\begin{center}
\includegraphics[width=16.0cm]{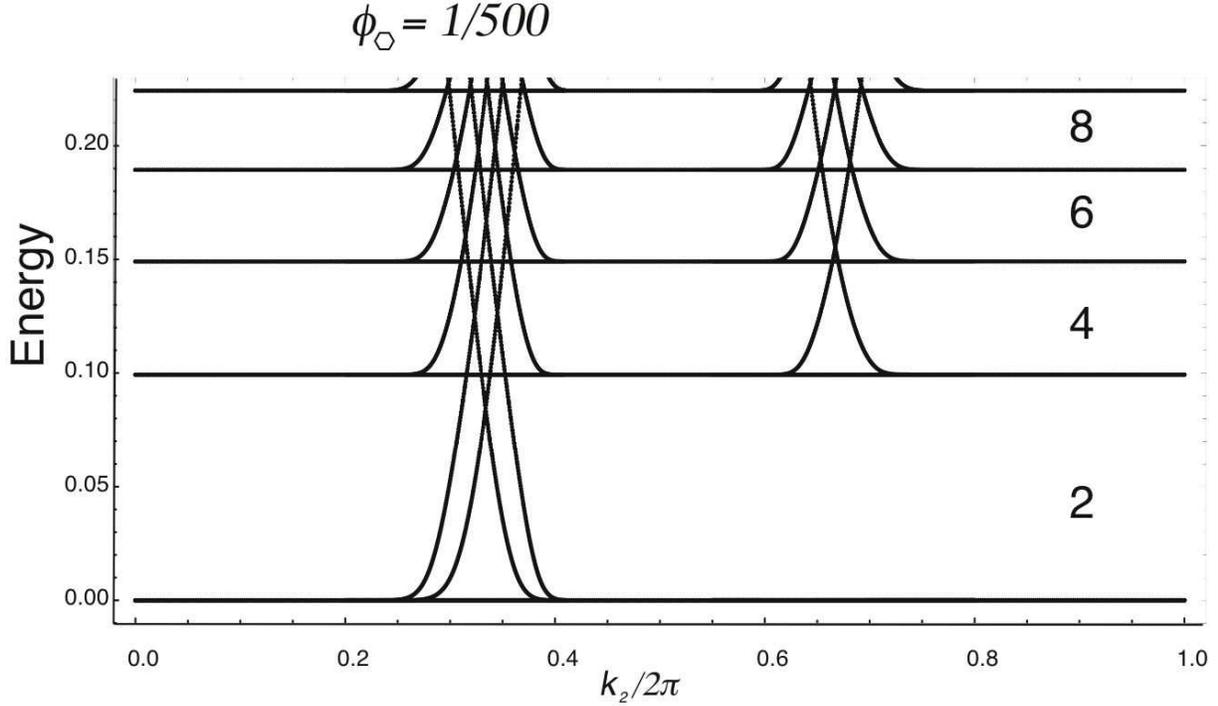} 
\end{center}
\caption{\label{label:edge}
Energy spectrum in the region  around the $n=0$ Landau level 
in the bilayer graphene with zigzag edges and a 
periodic boundaries orthogonal to those (along $e_2$).  
Here we have a relatively small magnetic field $\phi=1/500$, 
with $\gamma_0=1.0$ and $\gamma_1=0.2$. 
The attached numbers 
denote the Hall conductivity 
when $E_F$ is in each interval between the Landau levels, 
which coincide with the number of edge modes (curves 
traversing different Landau levels). 
}
\end{figure}

In Fig.\ref{label:edge}, we have shown the spectrum of the bilayer graphene
on a cylinder. 
Here the translational symmetry along $e_2$ 
makes the wave number $k_2$ a good quantum number, 
so that the dispersion against  $k_2$ is depicted.  
The ``bulk-edge correspondence",
which dictates bulk topological features such as
the quantized Hall conductivity has a 
one-to-one correspondence to  
edge properties,   
allows us to obtain 
the Hall conductivity of the
system by counting the number of edge modes.

\subsection{Zero magnetic field}
For zero magnetic field, we can make a Fourier transform 
along the $\bm{e}_1 $ 
as
\begin{eqnarray}
u_\alpha, d_\alpha, t_\alpha (k_1 ,k_2  ) =
 \int _{-\pi }^{\pi} \frac {dk_1}{2\pi} 
e^{ik_1 j_1} u_\alpha, d_\alpha, t_\alpha (j_1,k_2),
\end{eqnarray}
we have a fully Fourier-transformed Hamiltonian.  
For the bilayer we have
\begin{eqnarray}
{\cal H}^{\rm bilayer} = \int \frac {d^2k}{(2\pi)^2}  {\cal H}({k} )
\\
{\cal H}({k} ) = 
  \gamma _0
( 1
+
 e^{-ik_2}
+
 e^{-i k_1} )
(d_\bullet ^\dagger   d_\circ 
+u_\bullet ^\dagger   u_\circ )
\nonumber \\
\quad 
+\gamma _1 
u _ \bullet ^\dagger d_ \circ 
+ \gamma _3
(
e^{-i k_1} 
+
 e^{-ik_2}
+
e^{-i(k_1+k_2)}
)
 d_\bullet  ^\dagger u_ \circ  
+ {\rm h.c.}
\\
= 
(
u_\bullet ^\dagger  ,
d_\bullet ^\dagger  ,
u_\circ ^\dagger ,
d_\circ ^\dagger 
)
\bm{H}^L_{b} 
\mvecfour
{u_\bullet }
{d_\bullet }
{u_\circ }
{d_\circ}
\\
\bm{H}_b^L
 =
\mmat
{O}{\bm{D} }
{\bm{D} ^\dagger }{O},\quad
\bm{D}  = 
\mmat
{\Delta }
{\gamma _1}
{\gamma _3'}
{\Delta }
\end{eqnarray}
where
\begin{eqnarray}
\Delta(k)  = \gamma _0(1+e^{-i k_1}+e^{-i k_2}),
\\
\gamma _3' = \gamma _3(e^{-i k_1}+e^{-i k_2}+e^{-i k_1}e^{-i k_2}).
\end{eqnarray} 

For the trilayer, we have
\begin{eqnarray}
{\cal H}^{\rm trilayer} = \int \frac {d^2k}{(2\pi)^2}  {\cal H}({k} ),
\\
{\cal H}({k} ) = 
  \gamma _0
( 1
+
 e^{-ik_2}
+
 e^{-i k_1} )
(d_\bullet ^\dagger   d_\circ 
+u_\bullet ^\dagger   u_\circ 
+t_\bullet ^\dagger   t_\circ 
)
+
\\
 ({\rm for}\ ABC)\quad
+\gamma _1 
(t _ \bullet ^\dagger u_ \circ 
+u _ \bullet ^\dagger d_ \circ )
+ \gamma _3'
(u_\bullet   ^\dagger t_ \circ  
+d_\bullet  ^\dagger u_ \circ  
 )
+ {\rm h.c.}
\\
({\rm for}\ ABA)\quad +\gamma _1 
(u _ \bullet ^\dagger t_ \circ 
+u _ \bullet ^\dagger d_ \circ )
+ \gamma _3'
(
t_\bullet   ^\dagger u_ \circ  
+ d_\bullet   ^\dagger u_ \circ  )
+ {\rm h.c.}
\\
= 
(
t_\bullet ^\dagger  ,
u_\bullet ^\dagger  ,
d_\bullet ^\dagger  ,
t_\circ ^\dagger ,
u_\circ ^\dagger ,
d_\circ ^\dagger 
)
\bm{H}^L_{b} 
\mvecsix
{t_\bullet }
{u_\bullet }
{d_\bullet }
{t_\circ }
{u_\circ }
{d_\circ}
\\
\fl\qquad
\bm{H}_t^L
 =
\mmat
{O}{\bm{D}^{ABC,ABA} }
{{\bm{D}^{ABC,ABA}} ^\dagger }{O},
\\
\bm{D}^{ABC}  
=
\matthree
{\Delta}{\gamma _1}{ 0}
{\gamma _3'}{\Delta}{\gamma _1}
{0}{\gamma _3'}{\Delta},\quad
\bm{D}^{ABA}  =
\matthree
{\Delta}{\gamma _3'}{ 0}
{\gamma _1}{\Delta}{\gamma _1}
{0}{\gamma _3'}{\Delta}
\end{eqnarray}

Further, we can extend the above argument to consider the ABC stacked graphene with a gegeral number ($p$) of layers.   
In the present representation the Hamiltonian simplifies into
\begin{eqnarray} 
H_p=\mmat{O}{D_p}{D_p ^\dagger }{O},\quad
D_p = \Delta {\bf 1} + \gamma_1 J_p+ \gamma _3' \tilde {J}_p ,
\end{eqnarray} 
where ${\bf 1}$ is a $p$-dimensional unit matrix and 
$J_p=
\left [\begin{array}{ccc}
0 & 1 & 0\\
0 & 0 & \ddots \\
\vdots &  \ddots&\ddots
\end{array}
\right]
$.

\subsubsection{Low-energy Hamiltonians around K and K'}

The Dirac points  (K and K' points) for the monolayer graphene 
are specified by $e^{-ik_1}=\omega $ and $e^{-ik_2}=\omega ^{2} $, $\omega ^3=1, \omega \ne 1$, at which $\Delta$ vanishes along with 
$\gamma _3'\to 0$. 
Around  the Dirac points we can expand them as
\begin{eqnarray}
\Delta = - i\gamma _0 \omega \xi,\quad 
\gamma _3' = i \gamma _3 \omega ^2  \xi^*,
\quad \xi= \delta k_1 + \omega \delta k_2, 
\end{eqnarray} 
As for the multilayer graphene, 
gapless momenta are determined by $\det D $'s.
For the bilayer, it is given as
\begin{eqnarray}
\det D^{AB}= \Delta ^2-\gamma _1 \gamma _3'=
 - \gamma_0^2 \omega ^2\left( \xi^2 - \frac {i \gamma _1 \gamma _3}{\gamma _0^2}  \xi^*\right).
\end{eqnarray} 
This vanishes at four points, $\xi=0, \xi_0, \omega \xi_0, \omega ^2 \xi_0$, 
where $\xi_0= (\gamma _1 \gamma _3/\gamma _0^2) e^{i \pi/6 } $.  
Namely, each Dirac point proliferates into four.  

For the trilayer, it is 
\begin{eqnarray}
\det D^{ABC} =\det D^{ABA} =\Delta ^3-2 \Delta \gamma _1 \gamma _3'.
\end{eqnarray} 
This gives Dirac cones
at $\xi = \sqrt{2}\xi_0, \sqrt{2}\xi_0\omega, \sqrt{2}\xi_0\omega^2$ and double zero at $\xi=0$.  
One may discuss the case of $p-$layers as well.

\begin{figure}
\begin{center}
\includegraphics[width=4.7cm]{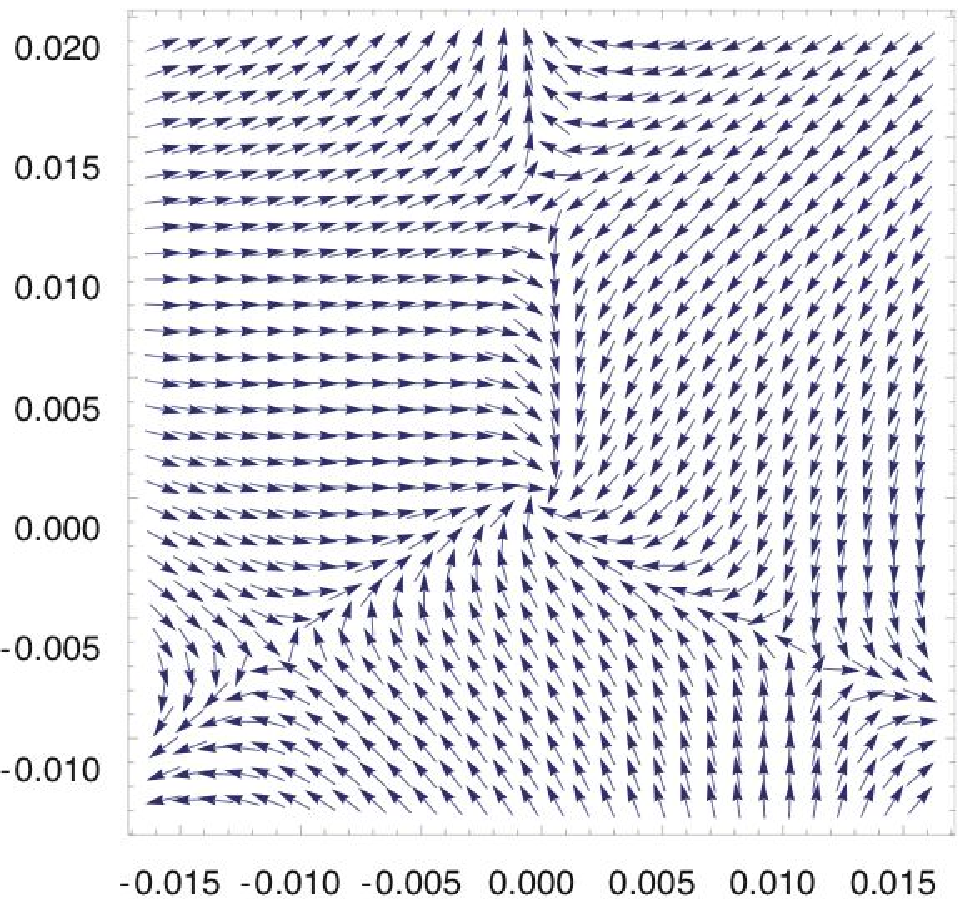} 
\includegraphics[width=4.7cm]{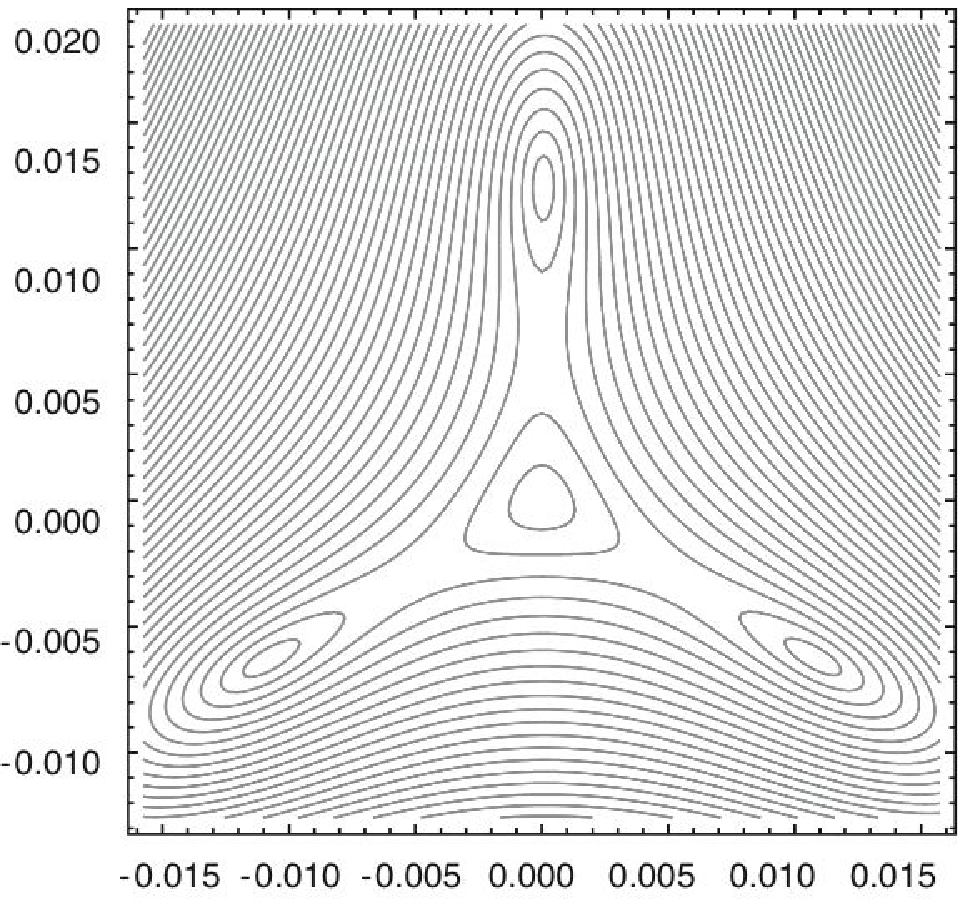}   
\includegraphics[width=5.5cm]{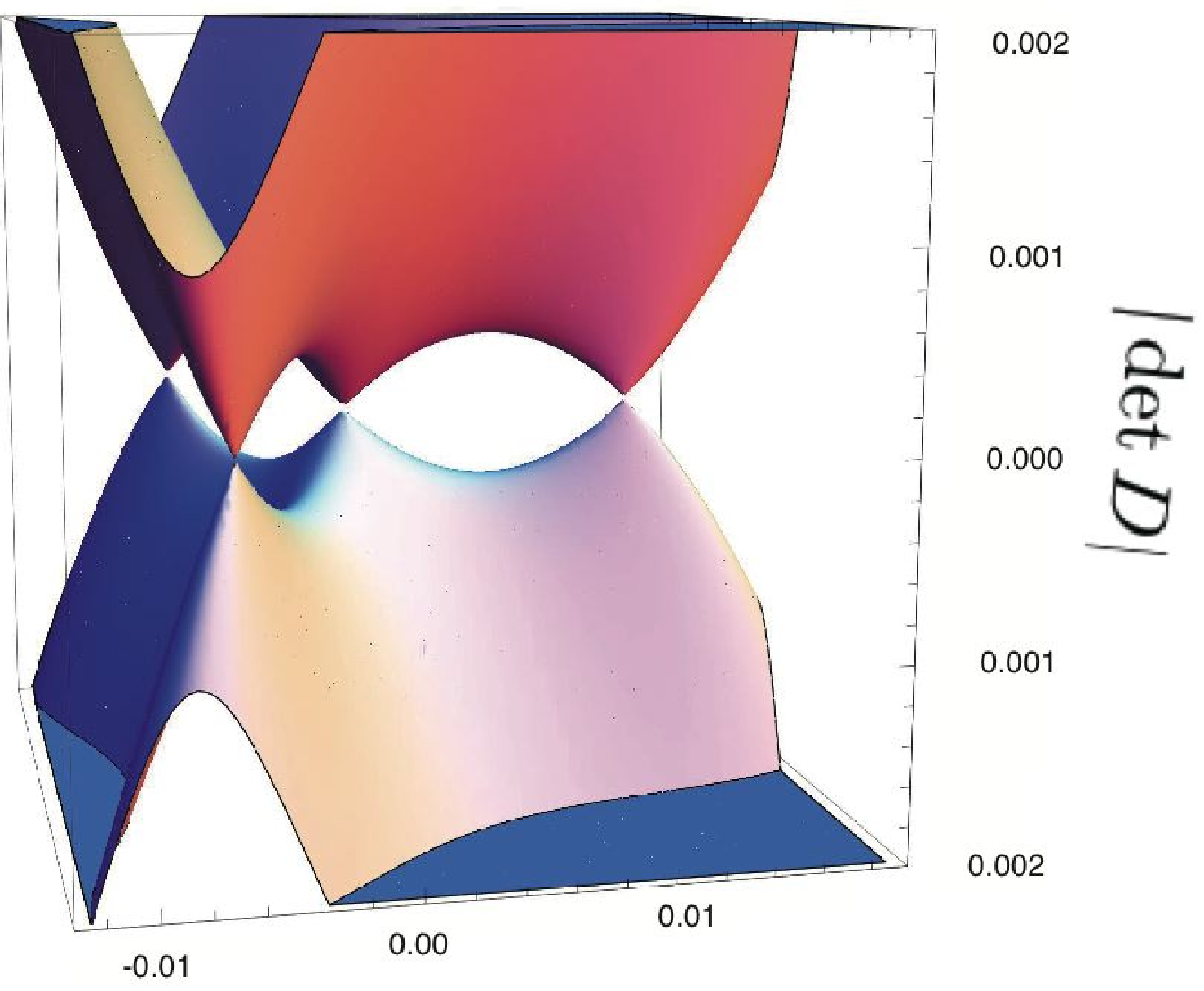}      
\end{center}
\caption{For the 
bilayer graphene 
in the presence of the trigonal warping ($\gamma_3\neq 0$), 
we show ${\rm Arg}\, \det D$ (left panel), 
contours for $|\det D|$ (center), and 
$\det D (\propto$ the band dispersion; right). 
Here we take $\gamma_0=3.1$,
$\gamma_1=0.4$, 
$\gamma_3=0.3$.
}
\label{label:D} 
\end{figure}

The Dirac points in the chiral-symmetric system at $E=0$ are 
specified by $\det D=0$, for which the low-energy dispersion 
is given by $\epsilon(k) \propto \pm |\det D|$.  
In Fig.\ref{label:D}, we have shown the low-energy dispersion 
along with $|\det D|$ and arg $\det D$ 
for a finite value of $\gamma _3$ 
in the bilayer system.  
From this plot we can see that the chiralities of the four 
Dirac points are +1, -1,-1,-1 or -1,1,1,1.

Now let us discuss the low-energy Hamiltonian 
in the absence of the trigonal warping ($\gamma _3=0$). 
If we denote the number of layers by $p$ 
in the ABC-stacked $p$-layered graphene (including bilayer), $D_p$
is triangular
\begin{eqnarray}
D_p (k)= \gamma _1
\left[
\begin{array}{cccc}
z(k) & 1 & 0 & \\
0 & z(k) & 1 &\ddots \\
 & \ddots & \ddots &\ddots 
\end{array}
\right],
\quad z(k) = \Delta(k) /\gamma _1,
\\
\det D_p (k)= \Delta ^p,
\end{eqnarray} 
where taking a continuous limit around K and K' points implies 
we are assuming $|z|\ll 1$. 
Precisely at the K and K' points, we have 
$D_p ^\dagger D_p= \gamma _1 ^2{\rm diag}\ (0,1,1,\cdots)$, 
i.e., there exists only one low-energy mode for $D_p ^\dagger D_p$ while the others have $\gamma _1^2$.
Then the chiral symmetry 
(i.e., $H_p$ is composed of $D_p$ as the off-diagonal block) 
implies that $H_p$ has two low-energy modes, 
$\pm \epsilon(k)$, with $\epsilon(k) = |\det D_p(k)|/\gamma _1^{p-1}
= \gamma _1|z(k)|^p$.
Here we can take 
the chiral basis introduced in previous sections 
to expand the low-energy doublet as
\begin{eqnarray}
\psi=(\psi_+,\psi_- )=
\mmat
{\psi_\bullet }{0}{0}{\psi_\circ },\quad 
\quad D_p D_p ^\dagger \psi_\bullet =\epsilon^2 \psi_\bullet ,
\quad D_p ^\dagger  D_p \psi_\circ  =\epsilon^2 \psi_\circ 
\label{eq:dddp}
\end{eqnarray} 
 $\psi_{\bullet, \circ }$ are 
 asymptotically normalized and
($|z|\ll 1$) given as
\begin{eqnarray}
\psi_\bullet   =
\mvecfour
{(-z^*)^{p-1} }
{\vdots}
{-z^*}
{1},\quad
\psi_\circ  =
\mvecfour
{1}
{-z}
{\vdots}
{(-z)^{p-1} }
\end{eqnarray} 
with $\psi_\bullet ^\dagger \psi_\bullet 
=\psi_\circ ^\dagger \psi _\circ =(1-|z|^p)(1-|z|^2) ^{-1} =1+{\cal O}(|z|^2)$.
They are consistent with Eq.(\ref{eq:dddp})
up to the errors arising from the  normalization of 
$\psi_\bullet ^\dagger \psi_\bullet
$ and 
$\psi_\circ ^\dagger \psi _\circ$.
Then projecting out the high energy sectors,
we have an effective Hamiltonian for the low-energy doublet 
formed 
by the chiral basis  $\psi$ 
(including the monolayer case) as
\begin{eqnarray}
H_p^{\rm eff}=\psi ^\dagger 
H_p
\psi
=
-\gamma _1\mmat
{0}{(-z)^p}
{(-z^*)^p}{0}
=
(-\gamma _1)^{-(p-1)}
\mmat
{0}{\Delta ^p}
{{\Delta^*} ^p}{0}
\end{eqnarray} 

As for the ABA trilayer, 
$(D^{ABA}) ^\dagger (D^{ABA}) $ has two 
low-energy modes around the K and K' points, 
which implies the low-energy modes of $H_3^{ABA}$ is spanned by
a 4-dimensional orthonormalized chiral basis as
\begin{eqnarray}
\fl\qquad \psi=(\psi_+,\psi_- )=
\left[
\begin{array}{cc}
\psi_\bullet  & 0 \\
0 & \psi_\circ 
\end{array}
\right]
=
\left[
\begin{array}{cccc}
{\psi_\bullet^{(1)} } &{\psi_\bullet^{(2)} }&{0}&{0}\\
{0}&{0}&{\psi_\circ^ {(1)} }&{\psi_\circ ^{(2)} } 
\end{array}
\right],\\
\quad \psi ^\dagger \psi= 
{\rm diag} ( 1,1,1,1)+{\cal O}(|z|),
\\
\fl\qquad 
D_p D_p ^\dagger \psi_\bullet^{(i)} =\epsilon_ i^2 \psi_\bullet^{(i)} ,
\quad D_p ^\dagger  D_p \psi_\circ^{(i)}  =\epsilon_i^2 \psi_\circ ^{(i)}
,\quad \epsilon_1  =  |\Delta |,\quad
\epsilon_2  =  \gamma _1 ^{-1} |\Delta |^2/\sqrt{2},
\end{eqnarray} 
where we have, up to the leading order, 
\begin{eqnarray}
\psi_\bullet^{(1)}  =
\frac {1}{\sqrt{2}}
 \mvecthree
{-1}
{0}
{1},
\quad
\psi_\bullet^{(2)} 
=
\frac {1}{\sqrt{2}}
\mvecthree
{1}
{-z^*}
{1},
\nonumber \\
\quad
\psi_\circ ^{(1)}  =
\frac {1}{\sqrt{2}}
 \mvecthree
{-1}
{0}
{1},
\quad
\psi_\circ ^{(2)} 
=
\frac {1}{2}
\mvecthree
{-z}
{2}
{-z}.
\end{eqnarray} 
Now we have
$
\psi ^\dagger H_3^{ABA} \psi 
=
\mmat{O}{\psi_\bullet ^\dagger D \psi_ \circ }
{\psi_\circ ^\dagger D ^\dagger \psi_\bullet }{O}
$
and 
$\psi_\bullet ^\dagger D \psi_ \circ  =\gamma _1\mmat{z}{0}{0}{-z^2/\sqrt{2}}$,
we end up with a simple decomposition, 
\begin{eqnarray}
H_3^{\rm ABA:eff}=\gamma _1\mmat{0}{z ^* }{z}{0}\oplus \frac {\gamma _1}{\sqrt{2}} 
\mmat{0}{(-z ^*)^2}{(-z)^2}{0}
= H_1^{\rm eff} \oplus H_2^{\rm eff}/\sqrt{2}.
\end{eqnarray} 

\subsubsection{Landau levels}

Around the Dirac cones, we can expand
the effective Hamiltonian, as in the monolayer case, 
as
\begin{eqnarray}
\fl\quad 
H_1\to H_1^{\rm eff}
=
(
\bm{\sigma } \cdot 
\bm{X}) \delta k_x
+
(\bm{\sigma } \cdot \bm{Y} 
)\delta k_y, \quad
X   = 
\mvecthree
{ {\rm Re}\Delta_x }
{-{\rm Im}\Delta_x }
{0},\quad 
Y   = 
 \mvecthree
{{\rm Re}\Delta _y}
{-{\rm Im}\Delta _y}
{0}
\end{eqnarray} 
where 
$
\Delta _\alpha =\frac {\partial \Delta  }{\partial k_\alpha  }\big|_{k=K,K'} 
$.
Then we have
\begin{eqnarray}
\fl
(H_1^{\rm eff})^2 =(\hbar c  \overline{\delta k})^2,\quad
\overline{\delta  k}^2 \equiv [\delta k_x,\delta k_y] \Xi
\mvec{\delta k_x}{\delta k_y},\ 
\Xi= \frac {1}{|X\times Y|} 
\mmat{X\cdot X}{Y\cdot X}{X\cdot Y}{Y\cdot Y}
\end{eqnarray} 
where
$c^2 = |X\times Y|/\hbar^2$ is 
an effective ``light velocity" and 
$\overline{\delta k}$
is the averaged momentum near the Dirac cones with 
$ \det \Xi=(|X|^2|Y|^2-|X\cdot Y|^2)/|X\times Y|^2=1$.
Note that 
\begin{eqnarray}
X\times Y =
\mvecthree{0}{0}{{\rm Im}\, \Delta _x  \Delta _y^*}
=\mvecthree{0}{0}{\chi \hbar ^2 c^2}
\end{eqnarray} 
where $\chi={\rm sgn}\,{\rm Im\,} \Delta _x  \Delta _y^*$ 
is
the chirality of the Dirac cones. 

In a magnetic field, the 
Hamiltonian $H_1 ^C$ in 
the continuum limit is obtained by replacing 
$\hbar \delta k_\alpha \to \pi_\alpha= 
p_\alpha -e A_\alpha =-i\hbar \partial _\alpha - eA_\alpha $
where
$
{[}\pi_x,\pi_y{]} 
=i\hbar eB=i(\hbar/\ell_B)^2$ with 
$\ell_B=\sqrt{\hbar/eB}$ ($eB>0$).
As for the ABC stacked $p$-layered 
graphene (including monolayer and bilayer), 
the effective continuum Hamiltonian is obtained as the extention of 
the McCann-Fal'ko \cite{McCann06} as
\begin{eqnarray}
H_p^{\rm eff} &\to & H_p^C \equiv  (- \gamma _1)^{-(p-1)}
\mmat{0}{(\Pi ^\dagger)^p }{\Pi^p}{0},
\label{ABC-p-layer-B}
\\
\Delta &\to& \Pi^\dagger \equiv \hbar ^{-1} (\Delta _x \pi_x+\Delta _y \pi_y).
\end{eqnarray} 
This is another derivation of the $p$-th layer effective Hamiltonian
discussed by Koshino-McCann\cite{KoshinoABC}
for the chiral symmetric case.
Here 
the commutation relation of $\Pi$ is given as
\begin{eqnarray}
[\Pi,\Pi ^\dagger ] =2\hbar ^{-2} {\rm Im\,} \Delta _x \Delta _y^*
= 2(c\hbar/\ell_B)^2 \chi.
\label{Pi-commutation}
\end{eqnarray} 
Then one can define a canonical boson operator $a$, ($[ a,a ^\dagger ]=1$) as
\begin{eqnarray}
\Pi 
 =\mchss
{a\,  (c\hbar/\ell_B)\sqrt{2}}{(\chi=+1)}
{a ^\dagger\,  (c\hbar/\ell_B)\sqrt{2}}{(\chi=-1)}.
\end{eqnarray} 

For the positive chirality $\chi=+1$,
one has
\begin{eqnarray}
\fl\qquad (H_p^C )^2= 
N_p^2
\mmat
{(a ^\dagger )^p a ^p}
{0}
{0}
{a ^p(a ^\dagger )^p }
=
 N_p ^2
\mmat
{\prod_{i=1}^p(n-i+1)}
{0}
{0}
{\prod_{i=1}^p(n+i)}
\end{eqnarray} 
where 
$N_p=\gamma _1(\gamma _1 ^{-1} c \sqrt{2e\hbar B})^{p}$, 
$n=a ^\dagger a $, and 
\begin{eqnarray}
(a ^\dagger )^2 a ^2 =a ^\dagger n a =(n-1)n,\
(a ^\dagger )^3 a ^3 
=(n-2)(n-1) n,\
\cdots
\\
a ^2 (a ^\dagger )^2
= (n+2)(n+1), \cdots.
\end{eqnarray}
Then we have a serise of non-equally spaced Landau levels as
\begin{equation} 
E_n = \pm N_p\sqrt{n(n-1)(n-2)\cdots(n-p+1)}.
\label{ABC-p-layer-LL}
\end{equation}
The $\chi=-1$ case follows trivially.


\subsection{Lifshitz transition in magnetic fields}
In this section, let us focus on the Lifshitz transition 
caused by the trigonal warping term $\gamma  _3$, 
by calculating the energy levels on a honeycomb lattice.
Using the chiral symmetry, we can diagonalize $D ^\dagger D$ within the
low-energy sector, 
 which suffices to discuss the Landau levels near $E=0$. 
Since $D$ is a sparse matrix, 
we can handle quite small flux $\varphi$ numerically 
using the lattice Hamiltonian. 

\begin{figure}
\begin{center}
\includegraphics[width=8.0cm]{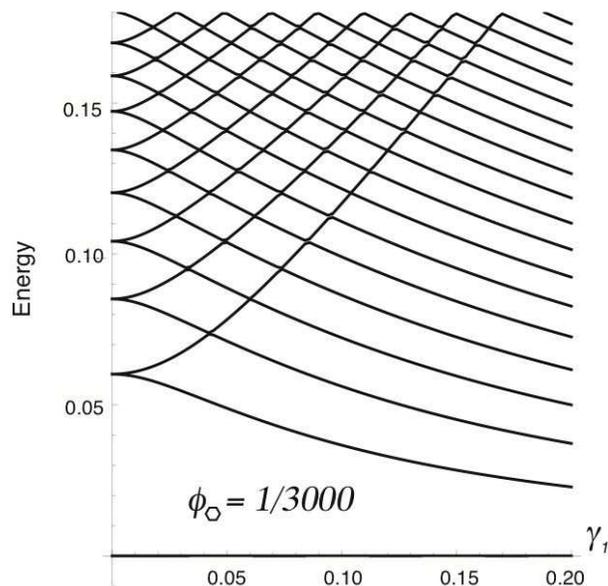} 
\end{center}
\caption{\label{label:gm1}
Low-energy Landau levels in the bilayer graphene 
for the magnetic field  $\phi=1/3000$ as a function of $\gamma_1 $
 (with $\gamma_0=1.0$ and $\gamma_3=0$ here).
}
\end{figure}

Let us first show the energy levels of the bilayer graphene
as a function of $\gamma _1$ in Fig.\ref{label:gm1}.
One can observe that 
larger $\gamma _1$ produces 
larger bonding-antibonding splitting, so that 
the low-energy sector is projected out from 
the high energy one. 

\begin{figure}
\begin{center}
\includegraphics[width=7.0cm]{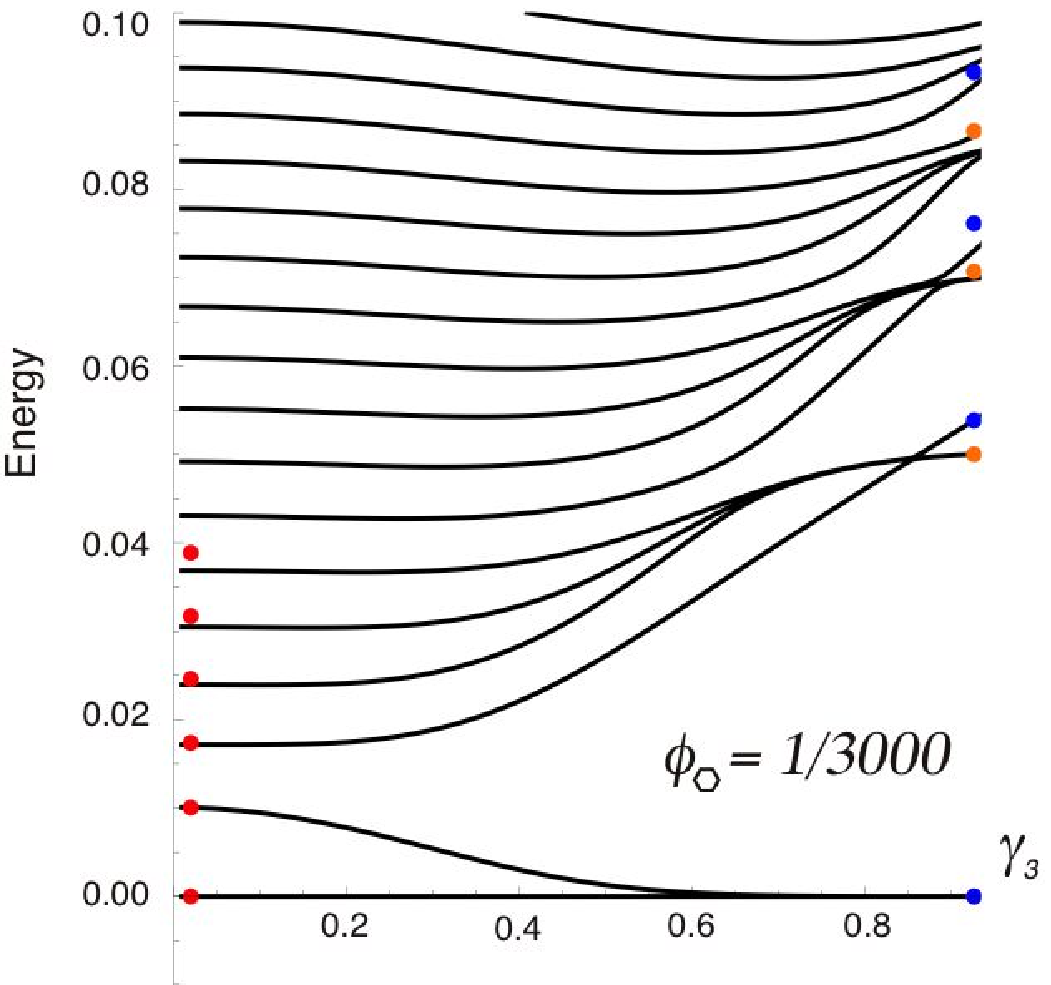} 
\includegraphics[width=8.5cm]{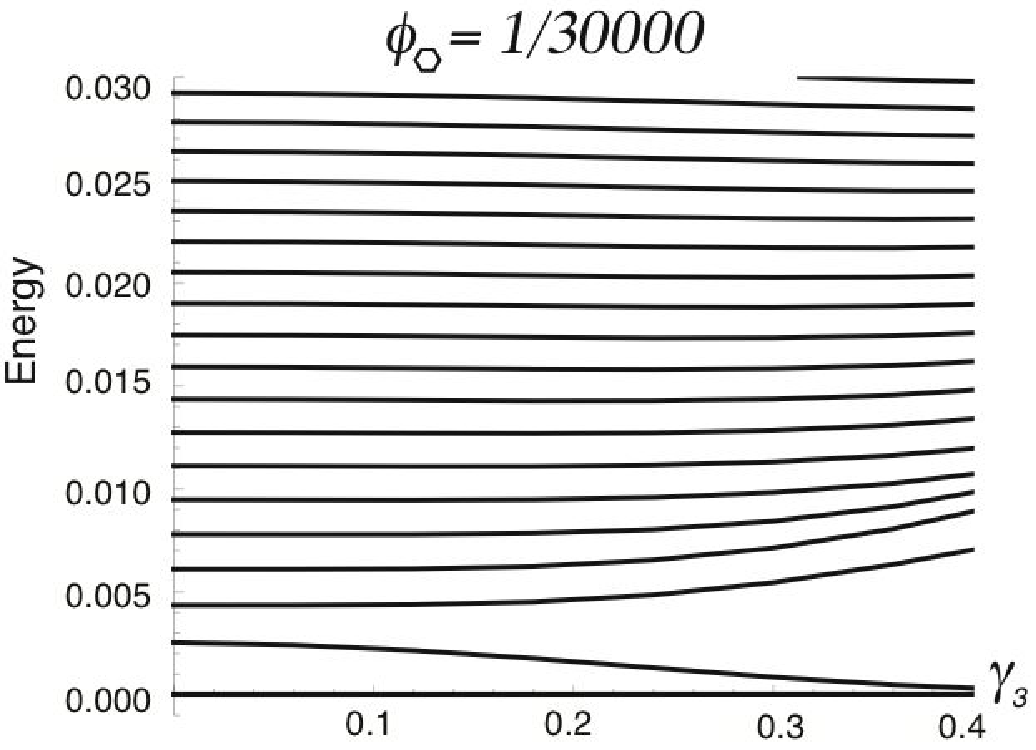} 
\end{center}
\caption{\label{label:gm3}
The Lifshitz transition caused by  $\gamma_3 $ is observed.
(left) Low-energy Landau levels in the bilayer graphene in a
magnetic field  $\phi=1/3000$
 ($\gamma_0=1.0$ and $\gamma_1=0.5$).
The red points at $\gamma _3=0$
denote a sequence $Const.\sqrt{n(n-1)}$, ($n=0,1,2,\cdots$).
We have also displayed the sequences $\propto \sqrt{n}$ for 
$n=0,1,2,3$ (blue dots) and for $n=1,2,3$ (orange).  
(right) The same plot for  realisic 
$\gamma _1/\gamma _0=0.2$ and much smaller flux $\phi=1/30000$.
}
\end{figure}

Figure \ref{label:gm3} , on the other hand, displyas the 
Landau levels as a function of the $\gamma _3$.  
 One can see that the Landau levels at
$\gamma _3=0$ are adiabatically connected to two classes of 
Landau levels at $\gamma _3\ne 0$, which correspond to those of 
two kinds of 4 Dirac cones 
with differente fermi velocity. 
To observe the phenomena at the  moderate value of the flux,
we take artificially large value for $\gamma _1/\gamma _0=0.5$
in the left pannel of the Figs.\ref{label:gm3}.
In the right pannel, the same plot for a 
much smaller, realistic $\gamma _1$ and the magnetic field.

\section{Optical responses in bilayer and trilayer graphene}
Let us move on to discuss optical responses in bilayer and trilayer graphene
in the quantum Hall regime.
We focus on the optical longitudinal conductivity $\sigma_{xx}(\omega)$,
which descrones the optical absorption,
and the optical Hall conductivity $\sigma_{xy}(\omega)$,
which is directly related to the Faraday rotation,
since the Faraday rotation angle $\Theta_H$ is proportional to  $\sigma_{xy}(\omega)$ in the quantum Hall regime as
$$
\Theta_H  \simeq
\frac{1}{(n_0+n_s)c_0\varepsilon_0}\sigma_{xy}(\omega),
$$
where $c_0$ is the speed of light and $n_0 (n_s)$ is the refractive index of the air (substrate)\cite{oconnell82}.

The Faraday rotation for monolayer graphene is studied theoretically \cite{gusynin-sxy,morimoto-opthall}, 
and experimentally starts to be measured \cite{crassee2010giant}.  
Further, there are growing interests in the optical properties 
of bilayer and trilayer graphene, 
given that their 
electronic structures are distinct from that of monolayer graphene.
As we have seen in Sec. 4,
bilayer graphene 
has two parabolic bands touching at the K and K' points \cite{nov-bilayer,mccann-falko}.  
Trilayer graphene comes with two different types of stacking, 
ABA and ABC, where the former is described effectively 
as a mixture of monolayer and bilayer bands, while the latter shows a cubic band as Eq.\ref{ABC-p-layer-B} with $p=3$.
Let us discuss optical responses for multilayer graphenes with these different lattice structures.  
Especially, we argue the effect of the presence or absence of 
the chiral symmetry gives a significantly different optical responses of $n=0$ LL 
in ABA trilayer graphene.

The optical longitudinal ($\sigma_{xx}(\omega)$) and Hall  ($\sigma_{xy}(\omega)$) conductivities are evaluated from the Kubo formula  as 
\begin{equation}
\sigma_{\alpha \beta}(\omega) =
 \frac{\hbar}{iL^2} \sum_{ab} j_\alpha^{ab} j_\beta^{ba} 
\frac{f(\epsilon_b) - f(\epsilon_a)}{\epsilon_b-\epsilon_a}
\frac{1}{\epsilon_b-\epsilon_a-\hbar\omega-i\eta},
\label{kuboformula}
\end{equation}
where $f(\varepsilon)$ is the Fermi distribution, 
$\epsilon_a$ eigenenergies, $\eta$ a small energy 
cutoff for a stability of the calculation, 
and $j_\alpha^{ab}$ is the matrix element of the current operator 
${\bm j} = \partial H/\partial {\bm A}$.

\subsection{Bilayer graphene}

\begin{figure*}[tb]
\begin{center}
\includegraphics[width=0.8\linewidth]{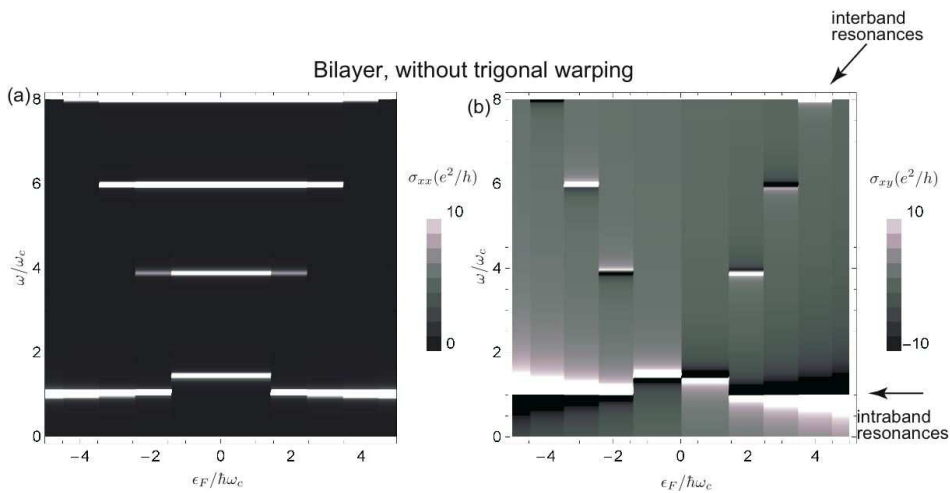}
\end{center}
\caption{For the bilayer graphene QHE system without the trigonal warping effect ($\gamma_3=0$), we show 
(a) optical longitudinal conductivity $\sigma_{xx}(\epsilon_F,\omega)$, and 
(b) optical Hall conductivity $\sigma_{xy}(\epsilon_F,\omega)$ plotted 
against the Fermi energy $\epsilon_F$ and the frequency $\omega$.
}
\label{wo-warping}
\end{figure*}

We start with calculating 
the optical Hall conductivity using Kubo formula Eqn.\ref{kuboformula} for the bilayer system without the trigonal warping effect (Eq.\ref{ABC-p-layer-B} with $p=2$). 
The result for the optical longitudinal and Hall conductivities are shown in Fig.\ref{wo-warping}(a,b).
If we label LLs with the Landau index $n$ and an electron/hole band index $s=\pm$,
an intra-band transition $(n,s) \to (n+1,s)$ occurs around 
$$
\omega_{\rm  intra} \sim  \omega_{\rm bilayer}=2 v^2 
e B/\gamma_1,
$$
with the Fermi velocity of the Dirac cone $v$ ($=c$ in Sec. 4), 
since LLs are almost equally spaced, 
unlike in the monolayer case where LL energy $\propto \sqrt{n}$ is not equally separated.  On top of this, 
there are inter-band transitions across the band-touching point, obeying a selection rule $(n,-s) \to (n+1,s)$.  
Thus the inter-band transition occurs around 
$$
\hbar\omega_{\rm inter} \simeq 2 |\epsilon_F|,
$$ 
for large enough $n$.

\subsection{ABA trilayer graphene}

For ABA stacked trilayer,
the effective Hamiltonian around $K_+$/$K_-$ points 
is given by a $6\times 6$ matrix 
(the dimension being 2 sublattices $\times$ 3 layers) 
as \cite{guinea-stacks06,partoens2006graphene,lu2006influence,koshino-ando07}
\begin{equation}
H_{\rm ABA}=
\left(
\begin{array}{cccccc}
0 &v \pi^\dagger & 0&v_3 \pi &\gamma_2/2&0\\
v \pi  &\Delta'& \gamma_1&0 &0 & \gamma_5/2 \\ 
0& \gamma_1&\Delta' & v \pi^\dagger &0 & \gamma_1\\ 
v_3 \pi^\dagger&0& v \pi  &0 &v_3 \pi^\dagger&0 \\ 
\gamma_2/2&0& 0&v_3 \pi &0& v \pi^\dagger  \\ 
0&\gamma_5/2& \gamma_1 &0& v \pi&\Delta'  
\end{array}
\right),
\label{Haba-full}
\end{equation}
with $\pi=\pi_x+i\pi_y$, a velocity $v_3$ associated with $\gamma_3$ as $v_3=v \gamma_3/\gamma_0$, 
$\Delta'$ the on-site energy difference between
the atoms with and without vertical bond $\gamma_1$,
and $\gamma_2 (\gamma_5)$ the 
next-nearest interlayer hoppings between $A1$ and $A3$ 
($B1$ and $B3$).  

\begin{figure*}[tb]
\begin{center}
\includegraphics[width=0.8\linewidth]{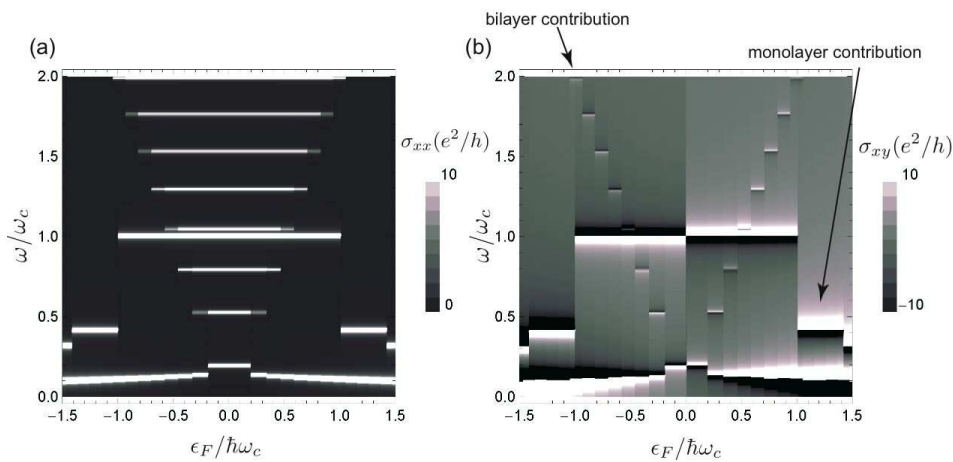}
\end{center}
\caption{For the ABA-stacked trilayer graphene QHE system 
(with $\gamma_1/\hbar \omega_c =5$) we show 
(a) optical longitudinal conductivity $\sigma_{xx}(\epsilon_F,\omega)$, 
and (b) optical Hall conductivity $\sigma_{xy}(\epsilon_F,\omega)$  plotted 
against the Fermi energy $\epsilon_F$ and frequency $\omega$.
}
\label{trilayer-aba}
\end{figure*}
First, we only retain $\gamma_0, \gamma_1$ terms in Eq.\ref{Haba-full}, 
in which case the Hamiltonian is chiral-symmetric.
In Fig.\ref{trilayer-aba}(a,b), the optical longitudinal and Hall conductivities $\sigma_{xx}(\epsilon_F,\omega), \sigma_{xy}(\epsilon_F,\omega)$
 are plotted against the Fermi energy $\epsilon_F$ and frequency $\omega$.  
The magnetic field is chosen so that an interlayer hopping energy is $\gamma_1/\hbar \omega_c =5$, with the monolayer cyclotron energy $\omega_c= \sqrt{2} v/\ell=v \sqrt{2eB/\hbar}$.
We can see a monolayer contribution (Dirac cyclotron frequency $=  \omega_c$) and a  bilayer contribution with a cyclotron energy,
$
\hbar\omega_{\rm{bilayer}}/ \sqrt{2}
,
$
both of which show intra-band and inter-band transitions.
For moderate magnetic fields $B \sim 1$ T, cyclotron frequency for monolayer is much larger than that for bilayer  $\omega_c \gg \omega_{\rm{bilayer}}$.
In this case, 
$\sigma_{xx}$ is an even function with respect to $\epsilon_F=0$, while 
$\sigma_{xy}$ is odd due to the electron-hole symmetry, a consequence of the chiral symmetry.
The jump of $\sigma_{xy}$ at $\epsilon_F=0$ is related to the chiral zero modes.

\begin{figure*}[tb]
\begin{center}
\includegraphics[width=0.95\linewidth]{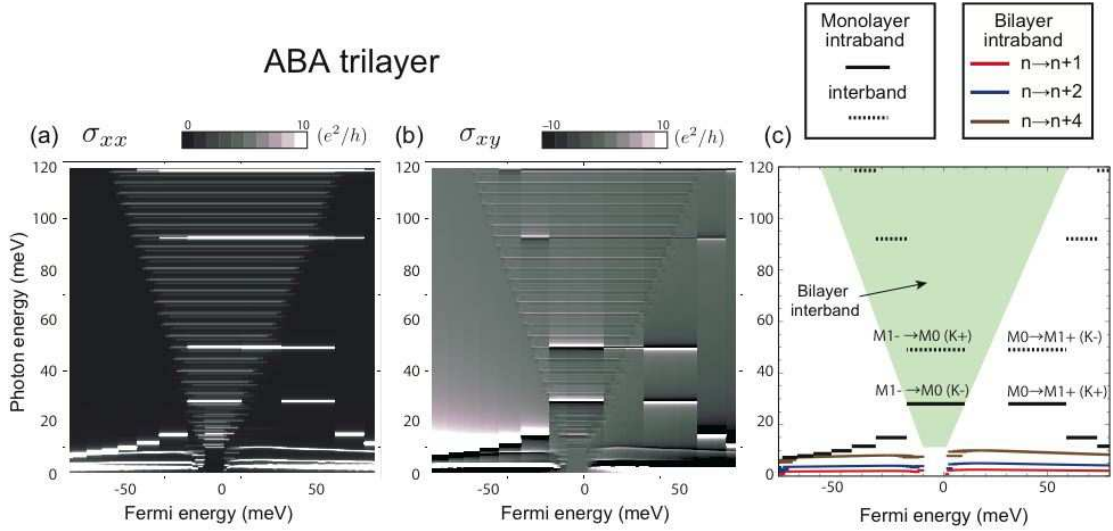}
\end{center}
\caption{
(a)
 Longitudinal $\sigma_{xx}(\epsilon_F,\omega)$ and 
(b)Hall 
$\sigma_{xy}(\epsilon_F,\omega)$ 
plotted against the Fermi energy $\epsilon_F$ and the frequency $\omega$
for a magnetic field $B=1T$. 
(c)A diagram indicating allowed resonances in $\sigma_{xy}$.  
}
\label{Fig-aba-full}
\end{figure*}

Now, if we include all the hopping terms in Eq.\ref{Haba-full},
the chiral symmetry is broken, and
we no longer have the chiral protection for zero modes (three zero energy LLs).  
We have then a massive Dirac band plus gapped bilayer bands with energy shifts.
As for hopping parameters,
we adopt the values for graphite,
$\gamma_0= 3.2$ eV,  $\gamma_1= 0.39$ eV,  $\gamma_3= 0.32$ eV,
$\gamma_2$=-0.020eV, 
$\gamma_5$=0.038eV, 
$\Delta'$=0.050eV 
\cite{graphite-abinitio,dresselhaus-graphite02}. 
Figures \ref{trilayer-aba}(a,b) depict the result for the optical longitudinal $\sigma_{xx}(\epsilon_F,\omega)$ and Hall conductivity $\sigma_{xy}(\epsilon_F,\omega)$
plotted against the Fermi energy $\epsilon_F$ and frequency $\omega$
for ABA stacked trilayer graphene QHE system.
We see contributions from monolayer-like Dirac LLs (labeled with M) and 
those from bilayer LLs (B), both of which exhibit intra-band and inter-band transitions.
Since Dirac cone is massive 
due to the chiral symmetry breaking 
and the zero-energy LL for the monolayer band (M0) is situated at the  bottom of
the conduction band for $K_+$ valley and the top of the valence band for $K_-$, 
$M0 \to M(1,+)$ resonance occurs at an energy lower than $M(1,-) \to M0$ for $K_+$, and vice versa for $K_-$.  
A cancellation of resonances in $\sigma_{xy}$, due to the opposite signs in current matrices, 
occurs between $M(1,-) \to M0$
for $K_+$ and  $M0 \to M(1,+)$ for $K_-$ 
for a region of Fermi energy between $M0(K_+)$ and $M0(K_-)$,
while this is not the case with $\sigma_{xx}$.  
For bilayer contributions, satellites appear 
as $B(n,\pm) \leftrightarrow B(n+1+3m,\pm)$ and $B(n,\pm) \leftrightarrow B(n+2+3m,\pm)$,
 since the trigonal warping term mixes $n$ LLs and $n+3$ LLs \cite{abergel-falko}.
The resonance frequency for intra-band transition within the conduction band is larger than those within the valence band, which is a consequence of 
the electron-hole asymmetry in the bilayer bands, triggered by a breaking of the chiral symmetry.  
A deviation in the  cyclotron mass for electron and hole bands prevents a complete cancellation between $B(n,-) \to B(n+1,+)$ and $B(n+1,-) \to B(n,+)$ transitions, which results in small interband transitions in a wide region of Fermi energy.

\subsection{ABC trilayer graphene}

The low energy effective Hamiltonian of ABC stacked trilayer graphene 
in a 2 by 2 matrix Eq.\ref{ABC-p-layer-B}
is a cubic form in the momentum,
if we neglect hopping terms but $\gamma_0, \gamma_1$.  
For this effective Hamiltonian, LL energy (Eq.\ref{ABC-p-layer-LL}) shows a magnetic-field dependence $\propto B^{\frac 3 2}$,
resulting in a smaller LL spacing compared to the single-layer LL $\propto B^{\frac 1 2}$ 
and bilayer LL $\propto B$ for weak magnetic fields.

\begin{figure*}[tb]
\begin{center}
\includegraphics[width=0.8\linewidth]{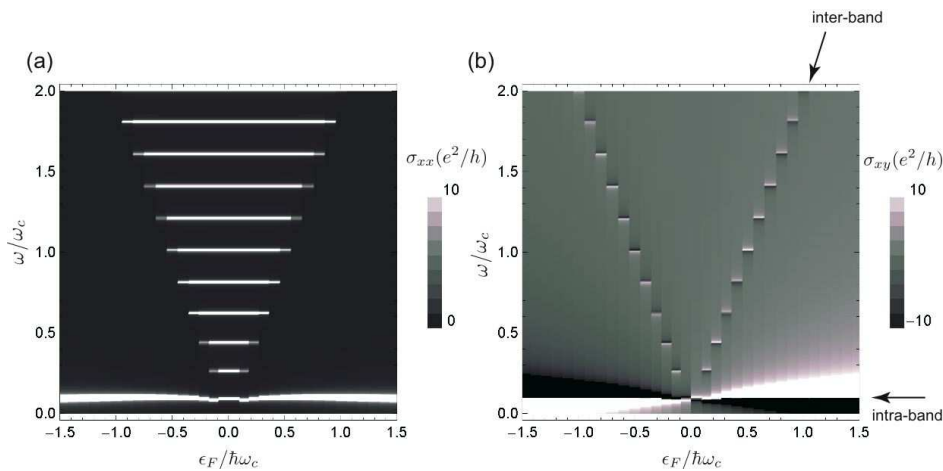}
\end{center}
\caption{For ABC-stacked trilayer graphene QHE system 
(with $\gamma_1/\hbar \omega_c =5$) we show 
(a) optical longitudinal conductivity $\sigma_{xx}(\epsilon_F,\omega)$, 
and (b) optical Hall conductivity $\sigma_{xy}(\epsilon_F,\omega)$  plotted 
against the Fermi energy $\epsilon_F$ and frequency $\omega$.
}
\label{trilayer-abc}
\end{figure*}

Now we turn to a result for the optical conductivities in ABC trilayer for an interlayer hopping energy $\gamma_1/\hbar \omega_c =5$, with monolayer cyclotron frequency $\omega_c= v \sqrt{2eB/\hbar}$.
Figures \ref{trilayer-abc}(a,b) show the optical longitudinal and  Hall conductivity $\sigma_{xy}(\epsilon_F,\omega)$
plotted against the Fermi energy $\epsilon_F$ and frequency $\omega$
for ABC stacked trilayer graphene QHE system.
In contrast to ABA stacking (Fig.\ref{trilayer-aba}),  we only see a single sequence of intra-band and inter-band transitions with a much smaller cyclotron energy than $\hbar \omega_c$ due to the dependence on magnetic fields $\propto B^{\frac 3 2}$.  
The inter-band transition occurs at $\hbar\omega \sim 2 \epsilon_F$ for the same reason as in the bilayer,
while the intra-band transition energy grows with increasing LL index $n$ as 
$$
\omega \sim (n+1)^{3/2}-n^{3/2} \sim n^{\frac 1 2},
$$
which explains why the intra-band resonance energy increases with $n$ in Fig.\ref{trilayer-abc}(d),
while it decreases with $n$ in the case of the monolayer graphene \cite{morimoto-opthall}.  
We can then predict in general that,  for ABC $p$-layered graphene, the intra-band transition occurs at 
$\omega\sim n^{p/2-1}$, and the inter-band transition at $\omega\sim 2\epsilon_F$.
Thus we end up with the intra-band transition energies 
that exhibit different behaviors with Landau index $n$ for monolayer ($\propto n^{-1/2}$), bilayer (constant) and ABC trilayer ($\propto n^{1/2}$) graphenes,
while the inter-band transition energies are qualitatively the same with $\sim 2\epsilon_F$.




\section*{Acknowledgement}
We thank Mikito Koshino for fruitful discussions.
The work is supported in part by Grants-in-Aid for Scientific 
Research, No.23340112 and No.23654128 from JSPS.
The computation in this work has
been done in part with the facilities of the Supercomputer Center, Institute
for Solid State Physics, University of Tokyo.

%
\section*{References}

\providecommand{\newblock}{}

\end{document}